\documentclass[a4paper,reqno,superscriptaddress,nofootinbib, pra, aps,10pt]{revtex4-2}
\usepackage[centertags]{amsmath}
\usepackage{amsfonts}
\usepackage{amssymb}
\usepackage{amsthm}
\usepackage{titlesec}
\usepackage{newlfont}
\usepackage{stmaryrd}
\usepackage{mathrsfs}
\usepackage{mathtools}
\usepackage{euscript}
\usepackage{graphicx}
\usepackage{enumerate}
\usepackage{todonotes}
\usepackage{color}
\usepackage{float}
\usepackage{orcidlink}
\usepackage{subfigure}
\usepackage{caption}
\usepackage{subcaption}
\def\la{\langle}
\def\ra{\rangle}

\usepackage{tikz}
\usepackage{pgf}
\usetikzlibrary{positioning,fit,calc}
\usetikzlibrary{arrows,automata}
\usepackage{wrapfig}
\usepackage{amscd}
\usepackage{cancel}
\usepackage{hhline}
\usepackage{import}
\usepackage{changes}
\date{\today}
\let\oldsqrt\sqrt
\def\sqrt{\mathpalette\DHLhksqrt}
\def\DHLhksqrt#1#2{
	\setbox0=\hbox{$#1\oldsqrt{#2\,}$}\dimen0=\ht0
	\advance\dimen0-0.2\ht0
	\setbox2=\hbox{\vrule height\ht0 depth -\dimen0}
	{\box0\lower0.4pt\box2}}

\newcommand{\id}{\textrm{d}}
\let\ve=\varepsilon 
\usetikzlibrary{3d,decorations.markings}
\usetikzlibrary{decorations.pathmorphing}

\usepackage{xcolor,soul}

\begin{document}

\title{An agitated oscillator chain}

\affiliation{Department of Physics and Astronomy, KU Leuven}

\author{Aaron Beyen \orcidlink{0000-0002-4341-7661}}
\affiliation{Department of Physics and Astronomy, KU Leuven}

\author{Christian Maes \orcidlink{0000-0002-0188-697X}}
\affiliation{Department of Physics and Astronomy, KU Leuven}

\author{Ion Santra \orcidlink{0000-0002-9772-2880}}
\affiliation{Department of Physics and Astronomy, KU Leuven}

\begin{abstract}
\color{black} Reduced dynamical descriptions for spatially extended systems in contact with a nonequilibrium medium give essential information about extensions of fluctuation-dissipation relations and, at the same time, are relevant for many natural systems.  Here we study a chain of harmonic oscillators coupled to fast run-and-tumble particles, discretizing earlier work on an elastic string. We derive the effective Langevin dynamics, including the explicit discrete counterparts for the nonequilibrium-induced streaming term, friction coefficient, and noise amplitude. At sufficiently high bath persistence, the linear friction becomes negative, destabilizing the chain dynamics. However, numerical simulations confirm that the anti-damping is eventually arrested and results in a nonlinear stationary condition, whose properties are the central topic of this work. We characterize the long-time phase-space trajectories and their Rayleigh-like structure, identify multistage relaxation of the mean-squared displacement and mean-squared velocity, and establish a non-monotonic dependence of stationary fluctuations on bath persistence. We further analyze the stationary distributions of the displacement and velocity, as well as the spatial correlations, which develop damped oscillations in the strongly persistent regime. The transfer of activity thus transforms a passive harmonic chain into a self-sustained fluctuating medium with many-body Rayleigh-like dynamics, pulsating displacements, and persistent velocities.
\end{abstract}

\maketitle

\section{Introduction}
Membranes, surfaces, elastic substrates, and even spacetime manifolds are generally subject to (bending) fluctuations \cite{brownianstring,Nelson2004Membranes,Monzel2016ShapeFluctuations, Rovelli_2013, structurefunctiondynamics}, which may, for instance, manifest themselves as ripples and waves \cite{Cerda2003Wrinkling, surfacewaves}. The origin of these fluctuations and deformations can vary, from interactions with surrounding thermal particles and heat baths to variable mechanical stresses in mechanobiology, \cite{Wang2006Mechanobiology}, and even quantum effects, \cite{stochasticgravity}. However, the Physics of Life \cite{NationalAcademies2022PhysicsOfLife,activematterbook} has unveiled new types of fluctuations, carrying the life signature, with novel mechanisms by which their stochastic motion is transferred to other systems \cite{pei2024inducedfrictionprobemoving, pei2025transferactivemotionmedium, beyen2025couplingelasticstringactive}. In that context, an interesting challenge is to understand how an elastic chain or string (the system) is influenced by coupling to an active matter bath, \cite{beyen2025couplingelasticstringactive, king2026activeparticlesdestabilizepassive, Grover_2025, davideactivemembrane}, and whether the chain becomes ``active'' itself. These questions are relevant, \textit{e.g.} for the physics of polymers in an active medium \cite{polymeractivemedium, Anderson2022-ug, Kaiser_2014}, interface growth \cite{Cagnetta_2019,Bisht_2019, Jana_2024}, and applications in macroscopic active matter \cite{diskagainstmembrane,robots1, robots2}.  These systems also represent a more tractable test case before entering the more complicated world of higher-dimensional and curved branes agitated by large and active molecules.\\

In contrast with the majority of approaches that `construct active field theories' \color{black}\cite{cates2019activefieldtheories, teVrugt2025, vrugt2022derivepredictiveactivefield, whatisactivematter} \color{black}, in the present paper, we aim for a bottom-up description where we start by describing the coupling between the bath particles and an oscillator chain, which belongs to the general context of field-particle interactions \cite{beyen2025couplingelasticstringactive, Gambassi1, Gambassi2}. A simple choice is to use
an interaction energy linear in the oscillator displacement with a coupling strength that depends on the relative position of the bath particle and the (fixed) lattice site of the oscillator. Importantly, the coupling allows feedback where the bending fluctuations impact the particle dynamics, \cite{diffusiononmembrane, Gopalakrishnan_2004, Bialus_Rallabandi_Oppenheimer_2025}, which is essential for understanding the induced friction caused by the bath on the oscillator chain.\\
While thermodynamically consistent, the active or nonequilibrium aspect of the bath violates the equilibrium Einstein relation between noise and friction \cite{maesresponse, Baiesi2011Modified, activethibaut,Santra_2023, pei2024inducedfrictionprobemoving,sarkar2025tracer}. For systems coupled to active media, such violations can generate effective anti-damping and linear instabilities~\cite{pei2025transferactivemotionmedium,beyen2025couplingelasticstringactive}. Here, this mechanism is derived explicitly for a periodic harmonic oscillator chain, weakly coupled to an active bath of run-and-tumble particles (RTPs), \cite{rtp_distribution, rtpintertialspring, Tailleur_2008, ionfirst}. The latter move on a ring, allowing for the presence of a persistent current, while the oscillators' displacements are orthogonal to the ring. \\

We consider fast and persistent bath particles, \textit{i.e.} strongly out of equilibrium, while the heavy, hence slow, oscillator chain is passive and Newtonian. Assuming a separation between the slow chain dynamics and the fast active bath, we derive a reduced, \color{black} Markovian, \color{black} Langevin equation for the chain characterized by a streaming force, a friction kernel, and random noise, \cite{maesresponse, zwanzig, Schilling, Tanogami2022}. Due to the nonequilibrium driving, the friction can become negative when the bath particles are sufficiently persistent, leading to a linear instability in the small-velocity regime. This instability causes the velocity and displacement fluctuations to grow, eventually violating the assumed time-scale separation assumption. To understand the fate of the linear instability, we take recourse to numerical simulations of the full microscopic dynamics and find that, at late times, the chain reaches a nonequilibrium stationary state with slow relaxation, broad, non-Gaussian stationary distributions, and a limit-cycle structure in phase space. That behaviour is reminiscent of (a collection of) Rayleigh oscillators \cite{Rayleigh01041883,Chen_1994, quantumrayleigh}, where negative linear friction at small velocities is balanced by positive nonlinear damping at larger velocities. In other words, the activity of the bath has transformed the initial passive chain into a self-sustained fluctuating medium with Rayleigh-like dynamics, which is an example of the emergence or transfer of activity. Along the chain, the steady state develops spatial correlations with damped oscillatory structure, and the displacement-momentum cross-correlations become antisymmetric, indicating persistent nonequilibrium currents in the oscillator variables. \\
\color{black} While many active and biological systems are naturally described in the overdamped (high friction) limit due to a low Reynolds number, our primary goal is to derive reduced dynamics from a microscopic system--bath model. In that context, it is natural to start from a conservative (Newtonian) probe dynamics and subsequently integrate out the nonequilibrium bath degrees of freedom. Indeed, that is also how a fundamental derivation of Brownian motion starts, often by applying the standard tools for deriving reduced dynamics, \textit{e.g.}, projection operators and singular perturbation theory. At any rate, the overdamped limit is known to be subtle and, in some situations, is best taken \emph{after} deriving the effective dynamics rather than at the level of the microscopic equations. It does not mean that the larger (thermal) environment does not add friction and noise as well, but those are left unspecified here. \color{black} \\

\color{black} This manuscript continues our earlier work \cite{beyen2025couplingelasticstringactive} on a string coupled to an active bath. In particular, to make the paper self-contained, we adapt the same reduction framework to a discrete oscillator chain, yielding the discrete counterparts of the bath-induced streaming term, (negative) friction, and noise amplitude. In \cite{beyen2025couplingelasticstringactive}, the numerical simulations verified the predicted instability due to negative friction and showed that the string does not keep growing forever, \textit{i.e.} a saturation regime is expected. The main contribution of the present manuscript then lies in the quantitative analysis of this long-time dynamics. New elements include the long-time phase-space trajectories and their Rayleigh-like structure, the multistage relaxation of the mean-squared displacement and velocity, and the non-monotonic dependence of the stationary fluctuations on bath persistence. We further characterise the non-Gaussian stationary displacement and velocity distributions and the spatial correlations, including their damped oscillatory structure in the strongly persistent regime. \\
The discrete formulation also provides a natural connection with spring-based descriptions of elastic systems, including simplified bead--spring models of polymers. It allows us to study site-resolved observables and to compare their behaviour with the equilibrium harmonic chain. Furthermore, the finite chain is precisely the model integrated in the numerical simulations, so that comparing the analytical and numerical descriptions does not require an additional spatial discretisation of a continuum field. \color{black} \\

The rest of the paper is organized as follows. We start in Section \ref{section model} by presenting the dynamics of a periodic chain of confined oscillators that are weakly coupled to a bath of fast RTPs. The goal is to describe both the initial transient regime and the steady regime of a fluctuating oscillator chain. In Section \ref{section reduced dynamics}, we derive a linear instability at higher bath persistence where, due to a negative induced (linear) friction coefficient, the mean and variance of the oscillator displacements are growing in time. At late times, a steady or saturated regime of fluctuations is reached, which is the subject of Section \ref{section saturation regime}. We give the numerical evidence and quantitative analysis of this (now) ``active'' elastic chain.  The Appendix contains various clarifications and explicit computations, together with a guide for the numerical implementation. \\

\section{Setup}\label{section model}
We consider a chain of $n$ coupled oscillators with displacements $q_j(t)\in \mathbb{R}$, $j=0,\ldots,n-1$, arranged orthogonal to a ring of length $L$ with periodic boundary conditions $q_j = q_{j+n}$. The index $j$ labels the number of the oscillator, and we take $r_j = j L/n$ to be its lattice site on the ring. The chain is coupled to a bath of $N$ RTPs with positions $z_m(t)\in \mathbb{S}^1$, $m=1,\ldots, N$, moving on the same ring. Hence, the particles live on the circle, while the oscillators describe transverse degrees of freedom. The interaction between the chain and the bath is described by the potential\footnote{This follows the standard linear coupling in field theory between a height field $\phi(r,t)$ and an external (particle) force, current or source, \cite{Kardar_2007, qftcritical, padmanabhan, Gambassi1, Gambassi2}
\begin{align*}
    U_{\text{int}}([\phi],\color{black} \vec z \color{black}) = \sum_{m = 1}^N \oint \id r \ \phi(r,t) \ F(r-z_m(t))
\end{align*}
and is sometimes also called a monopole interaction \cite{Passegger_Verch_2025, Doukas_Lin_Hu_Mann_2013}. Due to its linearity, this potential is the simplest possible coupling between particles and scalar fields. Differently, the energy Eq.~\eqref{Uint} can be seen as the cross product in $(q_j-F(r_j-z_m))^2$ where the other terms can then be absorbed into confining potentials for the probe and bath.
}
\begin{align}\label{Uint}
U_{\text{int}}(\color{black} \vec q, \vec z \color{black}) = \sum_{m=1}^N \sum_{j=0}^{n-1} q_j(t)\, F(r_j - z_m(t))
\end{align}
\color{black} with $\vec q = (q_1,...,q_n)$ and $ \vec{z} = (z_1,...,z_N)$. \color{black} The function $F(r_j - z_m)$ is a (periodic) interaction kernel depending on the relative position along the ring between the lattice point of the $j$th oscillator and a bath particle located at $z_m$. We take $F(x)$ to be smooth and localized around $x=0$. \color{black} With this coupling, the oscillator chain can only interact with those bath particles that lie within the support of $F(x)$, such that the typical width of the kernel can be interpreted as the particle size. A delta function $F(x) = F_0 \ \delta(x)$ corresponds to exact point particles, but is also unbounded; thus it is preferred to have a function with finite spatial support, \textit{e.g.} a Gaussian interaction where the standard deviation represents the size of the bath particles. As the colloid particles move on the circle, one takes the circular equivalent of a Gaussian, for which the von Mises form is a \color{black} convenient choice, \cite{evans2000statistical},
\begin{align}\label{von mises}
F(x) = \frac{F_0}{I_0(p)} \exp\left[p \cos\left(\frac{2\pi}{L}x\right)\right] = F_0 \left[1 + 2 \sum_{k = 1}^\infty \frac{I_k(p)}{I_0(p)} \cos \left(\frac{2 \pi k}{L} x \right) \right]
\end{align}
where $I_k(p)$ is the modified Bessel function of order $k$. The parameter $p$ controls the localization: for $p\to\infty$, $F(x)$ approaches a periodic Dirac delta function (Dirac comb), while it is uniform for $p = 0$. This form of coupling has also recently been considered in the context of heat transport through oscillator chains, where it was shown to give rise to non-trivial transport phenomena \cite{krekels2026negative,gautama2026specific}. \\
Besides the coupling with the bath RTPs, there is also a potential energy for the oscillators, for which we take a quadratic function in the $q_j$, resulting in the harmonic chain,
\begin{align}\label{E(q)}
E(\color{black} \vec q \color{black}) = \sum_{j = 0}^{n-1}\left( \frac{\kappa}{2}(q_{j+1}-q_j)^2 + \frac{K}{2} q_j^2    \right)
\end{align}
with spring constants $\kappa, K$. \color{black} We added a local harmonic pinning term $\frac{K}{2} q^2$ since otherwise, due to the linear coupling in \eqref{Uint}, the chain lowers its energy upon reducing its height, implying an escape to minus infinity. To prevent that instability, we include the harmonic potential. \color{black} In conclusion, with forces derived from Eqs.~\eqref{Uint}, \eqref{E(q)}, we study the coupled dynamics\footnote{\color{black}  This dynamics is equivalent to the master equation for the density $\rho_{\text{tot}}(\vec q,\vec p,\vec z,\vec \sigma)$
\begin{align}
    \frac{\partial \rho_{\text{tot}}}{\partial t}(\vec q, \vec p,\vec z,\vec \sigma) &= \left(\cal L_{z,\sigma}^\dagger + \cal L_{q,p}^\dagger \right) \rho_{\text{tot}}(\vec q, \vec p, \vec z,\vec \sigma) \nonumber \\
  \cal L_{z,\sigma}^\dagger \rho_{\text{tot}}(\vec q, \vec p, \vec z,\vec \sigma) & = \sum_{m = 1}^N - \frac{\partial}{\partial z_m}\left[\left( v_0 \sigma_m + \mu_zf_q(z_m)\right) \rho_{\text{tot}}(\vec q, \vec p, \vec z,\vec \sigma) \right] + \alpha [\rho_{\text{tot}}(\vec q,\vec p,\vec z,- \sigma_m) - \rho_{\text{tot}}(q,p,\vec z, \sigma_m)] \label{generator Lzsigma}\\
   \cal L_{q,p}^\dagger \rho_{\text{tot}}(\vec q, \vec p, \vec z,\vec \sigma)  & =- \sum_{j=0}^{n-1} \frac{p_{j}}{M} \frac{\partial \rho_{\text{tot}}}{\partial q_{j}}(\vec q, \vec p, \vec z,\vec \sigma) + \left[\kappa (q_{j+1} + q_{j-1} - 2 q_{j})  - K q_{j} - \zeta  \,\sum_{m = 1}^N F(r_j-z_m)  \right] \frac{\partial \rho_{\text{tot}}}{\partial p_{j}}(\vec q, \vec p, \vec z,\vec \sigma) \nonumber 
\end{align}
with forward generators $\cal L_{z,\sigma}^\dagger, \cal  L_{q,p}^\dagger $ for the RTPs and oscillator chain respectively. The notation $\rho_{\text{tot}}(\vec q,\vec p,\vec z,- \sigma_m)$ implies that only the $m$th spin is flipped.
\color{black}}
\begin{subequations}
\begin{align}
M \frac{\id^2 q_j}{\id t^2}(t) &= \kappa \big(q_{j+1}(t) + q_{j-1}(t) - 2 q_j(t)\big) - K q_j(t) - \zeta \sum_{m=1}^N F(r_j - z_m(t)) \label{p dot original} \\
\frac{\id z_m}{\id t}(t) &= v_0 \sigma_m(t) - \zeta \mu_z \sum_{j=0}^{n-1} q_j(t)\, \partial_{z_m} F(r_j - z_m(t)) \label{original z dynamics}
\end{align}\label{eq:langevins0}
\end{subequations}
with coupling constant $\zeta > 0$, mobility $\mu_z$ and common propulsion speed $v_0$. We also introduced the independent two-state processes $\sigma_m(t)=\pm 1$ that flip at rate $\alpha$, defining a run-and-tumble dynamics with persistence time $\alpha^{-1}$ with autocorrelation function
\begin{align}\label{auto correlation}
     \left \langle v_0 \sigma_m(t) \ v_0 \sigma_{m'}(t') \right \rangle = \delta_{m, m'} \ v_0^2 e^{- 2 \alpha |t-t'|}
\end{align}

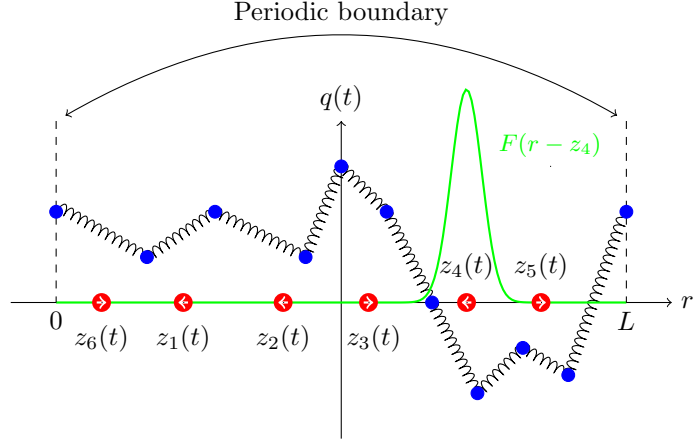
\begin{figure}[ht]
\centering
\begin{tikzpicture}[scale=1.20]
    % Axes
    \draw[->] (-3.14-0.5, 0) -- (3.14+0.5, 0) node[right] {$r$};
    \draw[->] (0, -1.5) -- (0, 2) node[above] {$q(t)$};

  % Curve made of springs connecting the discrete points
\draw[decorate, decoration={coil,aspect=0.3, segment length=3pt, amplitude=2pt}] 
    (-3.14,1) -- (-3.1,1.05);
\draw[decorate, decoration={coil,aspect=0.3, segment length=3pt, amplitude=2pt}] 
   (-3.1,1.05) -- (-3.14+1, 0.5);
%\draw[decorate, decoration={coil,aspect=0.3, segment length=3pt, amplitude=2pt}] 
 %   (-3,1.08) -- (-3.14+0.5, 0.75);
%\draw[decorate, decoration={coil,aspect=0.3, segment length=3pt, amplitude=2pt}] 
 %   (-3.14+0.5, 0.75) -- (-3.14+1, 0.5);
\draw[decorate, decoration={coil,aspect=0.3, segment length=3pt, amplitude=2pt}] 
    (-3.14+1, 0.5) -- (-3.14+1.75, 1);
\draw[decorate, decoration={coil,aspect=0.3, segment length=3pt, amplitude=2pt}] 
    (-3.14+1.75, 1) -- (-3.14/8,0.5);
\draw[decorate, decoration={coil,aspect=0.3, segment length=3pt, amplitude=2pt}] 
    (-3.14/8,0.5) -- (0, 1.5);
\draw[decorate, decoration={coil,aspect=0.3, segment length=3pt, amplitude=2pt}] 
    (0, 1.5) -- (0.5, 1);
\draw[decorate, decoration={coil,aspect=0.3, segment length=3pt, amplitude=2pt}] 
    (0.5, 1) -- (1, 0);
\draw[decorate, decoration={coil,aspect=0.3, segment length=3pt, amplitude=2pt}] 
    (1, 0) -- (1.5, -1);
\draw[decorate, decoration={coil,aspect=0.3, segment length=3pt, amplitude=2pt}] 
    (1.5, -1) -- (2,-0.5);
\draw[decorate, decoration={coil,aspect=0.3, segment length=3pt, amplitude=2pt}] 
    (2,-0.5) -- (2.5, -0.8);
\draw[decorate, decoration={coil,aspect=0.3, segment length=3pt, amplitude=2pt}] 
    (2.5, -0.8) -- (3.1,0.95);
\draw[decorate, decoration={coil,aspect=0.3, segment length=3pt, amplitude=2pt}] 
    (3.1,0.95) -- (3.14, 1);

    % particle position
  \def\zm{1.38}     % particle at x=0
  \def\sigma{0.17}  % width of Gaussian

  % Gaussian kernel
  \draw[green, thick, domain=-3.14:3.14, samples=200] 
    plot (\x, {1/(sqrt(2*pi)*\sigma) * exp(-(\x-\zm)^2/(2*\sigma^2))});

  % particle marker
  \fill[red] (\zm,0) circle (1.5pt);

  % axis
  \draw[] (2.2,1.5) -- (2.2,1.5) node[above, green] {\scriptsize $F(r-z_4)$};

    \draw[] (-0.6, 0.8) --  (-0.6, 0.8) node[above, blue] {\scriptsize $q_3(t)$};

    % Particles on the curve
    \filldraw[black] (-3.14+0.5, 0) circle (1pt);
    \filldraw[black] (-3.14+1.4, 0) circle (1pt);
    \filldraw[black] (-3.14+2.2,0) circle (1pt);
    \filldraw[black] (2.2, 0) circle (1pt);
    \filldraw[black] (1.38, 0) circle (1pt);
    \filldraw[black] (0.3, 0) circle (1pt);
     
     \node[circle, fill=red, inner sep=2.5pt] (A) at (-3.14+0.5, 0) {};
  \draw[white,->,thick] ($(A.center)+(-1.5pt,0)$) -- ($(A.center)+(1.5pt,0)$);

   \node[circle, fill=red, inner sep=2.5pt] (A) at (-3.14+1.4, 0) {};
  \draw[white,->,thick] ($(A.center)+(1.5pt,0)$) -- ($(A.center)+(-1.5pt,0)$);

   \node[circle, fill=red, inner sep=2.5pt] (A) at (-3.14+2.2,0) {};
  \draw[white,->,thick] ($(A.center)+(1.5pt,0)$) -- ($(A.center)+(-1.5pt,0)$);

   \node[circle, fill=red, inner sep=2.5pt] (A) at (2.2, 0) {};
  \draw[white,->,thick] ($(A.center)+(-1.5pt,0)$) -- ($(A.center)+(1.5pt,0)$);

  \node[circle, fill=red, inner sep=2.5pt] (A) at (1.38, 0) {};
  \draw[white,->,thick] ($(A.center)+(1.5pt,0)$) -- ($(A.center)+(-1.5pt,0)$);

  \node[circle, fill=red, inner sep=2.5pt] (A) at (0.3, 0) {};
  \draw[white,->,thick] ($(A.center)+(-1.5pt,0)$) -- ($(A.center)+(1.5pt,0)$);

    % Horizontal dashed lines for peaks and valleys
       \draw[dashed] (-3.14, 2) -- (-3.14, 0) node[below] {$0$};
     \draw[dashed] (-3.14+0.5, 0) -- (-3.14+0.5, 0) node[below, yshift=-5pt] {$z_6(t)$};
    \draw[dashed] (-3.14+1.4, 0)  -- (-3.14+1.4, 0)  node[below, yshift=-5pt] {$z_1(t)$};
     \draw[dashed] (-3.14+2.2,0) -- (-3.14+2.2, 0) node[below, yshift=-5pt] {$z_2(t)$};
      \draw[dashed] (0.3, 0) -- (0.3, 0) node[below, yshift=-5pt,  xshift=2pt] {$z_3(t)$};
    \draw[dashed] (1.38, 0) -- (1.38, 0) node[above, yshift=5pt] {$z_4(t)$};
    \draw[dashed] (2.2, 0)  -- (2.2, 0) node[above, yshift=5pt] {$z_5(t)$};
     \draw[dashed] (3.14, 2) -- (3.14, 0) node[below] {$L$};
     \draw[dashed] (-0.4, 0.7) -- (-0.4, 0) node[below, blue] {$r_3$};
     
     \node (l) at (-3.14,2) {};
     \node (r) at (3.14,2) {};
     \draw[<->] (l) to[bend left] node[above] {\text{Periodic boundary}} (r);
     
    \filldraw[blue] (-3.14,1) circle (2pt);
\filldraw[blue] (-3.14+1, 0.5) circle (2pt);
\filldraw[blue] (-3.14+1.75, 1) circle (2pt);
\filldraw[blue] (-3.14/8,0.5) circle (2pt);
\filldraw[blue] (0, 1.5) circle (2pt);
\filldraw[blue] (0.5, 1) circle (2pt);
\filldraw[blue] (1, 0) circle (2pt);
\filldraw[blue] (1.5, -1) circle (2pt);
\filldraw[blue] (2,-0.5) circle (2pt);
\filldraw[blue] (2.5, -0.8) circle (2pt);
\filldraw[blue] (3.14, 1) circle (2pt);
    \end{tikzpicture}
\caption{\raggedright A configuration of active particles (red disks) with associated spins (white arrow) on a ring, coupled to heavy harmonic oscillators (blue dots) through the force term $F(r-z_m)$ (green curve). The oscillators are connected through nearest-neighbor springs and satisfy periodic boundary conditions along the ring.}
\label{tikzpicturephi}
\end{figure}

That tumbling is the only stochastic ingredient in the dynamics of $(q,z)$. The interaction can be viewed as a sequence of localized kicks, where the active particles move along the ring and \color{black} interact with \color{black} the oscillator chain, exerting forces over a finite spatial region set by $F(x)$. The chain in turn exerts a reciprocal feedback on the particles through a coupling proportional to $\partial_x F(x)$. We recover the well-known thermal limit~\cite{santra2022universal} at temperature $T_{\text{eff}}$, for $v_0 \to \infty, \ \alpha \to \infty$ with $v_0^2/\alpha = 2\mu_z k_B T_{\text{eff}}$ constant, in which case the dichotomous noise emulates a Gaussian white noise. This limit inspires us to define
an effective inverse temperature $\beta_{\text{eff}} = 2 \alpha \mu_z/v_0^2$. \\
The dynamics of the total system is depicted in Fig.~\ref{tikzpicturephi}.

\section{Reduced dynamics}\label{section reduced dynamics}
We are interested in the reduced oscillator chain dynamics when integrating out the nonequilibrium bath. There exist standard techniques to do this, \textit{e.g.} projection operators \cite{Mori1, Mori2, zwanzig, Schilling, grabert1982projection}, nonequilibrium linear response theory and its pathspace action \cite{maesresponse, frenesy, over, under, Tanogami2022} or singular perturbation theory \cite{singularperturbationtheory, singularescaperate, Tanogami2022, Tikhonov1952, Lomov1992}. These methods require a time-scale separation in the sense that the mechanical variables $q_i(t), \  v_i(t)$ are slow compared to the active particles.  As explained in Appendix \ref{appendix time scale separation}, this is characterized by the small parameter $\varepsilon = \omega_{\text{chain}} L/v_0 \ll 1$ with $\omega_{\text{chain}} = \sqrt{\max\{\kappa, K\}/M}$ in the generator of the total system. Following these techniques, one obtains a reduced dynamics for the chain \color{black}(Itô calculus), \color{black}
\begin{align}
M \frac{\id^2 q_j}{\id t^2}
&= \kappa(q_{j+1}+q_{j-1}-2q_j)-Kq_j
-\zeta \bar{F}(q) -\sum_{\ell=0}^{n-1}\nu_{j\ell}(q)\frac{\id q_\ell}{\id t}
+\xi_j
\label{reduced dynamics 1}
\end{align}
\color{black} which is Markovian due to the assumed time-scale separation between the chain and bath particles, and is generically nonlinear in $q$. \color{black} The bath interaction has been decomposed into a mean force/streaming term $\bar{F}(q)$, an induced friction force with kernel $\nu_{j \ell }(q)$, and a fluctuating force $\xi_j$ \color{black} with zero mean and covariance 
\begin{align*}
    \left\langle \xi_j(t)\right\rangle &=0, \qquad \left\langle \xi_j(t) \ \xi_\ell(t')\right\rangle
=2B_{j\ell}(q)\ \delta(t-t')
\end{align*}
\color{black}
Following the standard projection operator formula \cite{pei2024inducedfrictionprobemoving, pei2025transferactivemotionmedium, beyen2025couplingelasticstringactive}, these induced terms are given by
\begin{align}
    \bar{F}(q) & = \sum_{m = 1}^N\left\langle F(r_j-z_m)\right\rangle_q^{\text{BO}} \nonumber \\
\nu_{j\ell}(q)
&= -\zeta \sum_{m = 1}^N \int_0^\infty \id\tau\,
\left\langle
F(r_j-z_m(\tau)) \, ; \, 
\frac{\partial \log \rho_q}{\partial q_\ell}(\vec z, \vec \sigma)
\right\rangle_q^{\text{BO}}
\nonumber \\
\color{black} B_{j\ell}(q) \color{black}
&\color{black} =\zeta^2 \sum_{m,m' = 1}^N \int_0^\infty \id\tau\,
\left\langle
F(r_j-z_m(\tau)) \, ; \, 
F(r_\ell-z_{m'}(0))
\right\rangle_q^{\text{BO}} \nonumber  \color{black}
\end{align}
\color{black} with $\vec{\sigma} = (\sigma_1,...,\sigma_N)$. \color{black}
Here, brackets $\left \langle \cdot \right \rangle$ without subscript indicate averages over the appropriate noise, while the average $\left \langle \cdot \right \rangle_q^{\text{BO}}$ and covariance $\left \langle \cdot \ ; \ \cdot  \right \rangle_q^{\text{BO}}$ are taken in the pinned, or Born-Oppenheimer (BO), ensemble $\rho_q(\color{black}\vec{z},\vec{\sigma}\color{black})$ of the active particles at fixed chain configuration $q$. \color{black} Using the forward generator $\mathcal L_{z,\sigma}^\dagger $ of the active particles at fixed $q$ in \eqref{generator Lzsigma}, the BO-distribution solves, \cite{rtp_distribution}, \color{black}
\begin{align}
0=\mathcal L_{z,\sigma}^\dagger \rho_q(\vec z, \vec \sigma)
= \sum_{m=1}^N
-\frac{\partial}{\partial z_m}
\left[
\left(v_0\sigma_m+\mu_z f_q(z_m)\right)\rho_q(\vec z,\vec \sigma)
\right]
+\alpha\left[
\rho_q(\vec z,-\sigma_m)-\rho_q(\vec z, \color{black}\vec \sigma \color{black})
\right]
\label{equation L rhoq = 0}
\end{align}
where $f_q(z)=-\zeta\sum_{j=0}^{n-1}q_j\,\partial_z F(r_j-z)$.
Then, for any observables $X(q,z,\sigma), Y(q,z,\sigma)$, we have 
\begin{align}
\left\langle X\right\rangle_q^{\text{BO}}
=
\sum_{\vec \sigma}\oint \id \vec z\,
X(q, \vec z, \vec \sigma)\ \rho_q(\vec z, \vec \sigma), \quad     \left \langle X  \ ; \ Y \right \rangle_q^{\text{BO}} = \left \langle \left( X - \left \langle X \right \rangle_q^{\text{BO}}\right)   \left( Y - \left \langle Y \right \rangle_q^{\text{BO}}\right) \right \rangle_q^{\text{BO}}\label{BO average mean force}
\end{align}
This distribution factorizes due to the independence of the $z-$particles $\rho_q(\vec z, \vec \sigma) = \prod_{m = 1}^N \rho^m_q(z_m, \sigma_m)$ and its solution was obtained in \cite{beyen2025couplingelasticstringactive, Bena_2003}. For completeness, we repeat the most important formulas of its solution in Appendix \ref{appendix rho_q}. Due to the independence of the bath particles, the mean force, friction coefficient, and noise covariance are of order $N$ and reduce to
\begin{align}
    \bar{F}(q) & = N\left\langle F(r_j-z)\right\rangle_q^{\text{BO}} \label{mean force} \\
\nu_{j\ell}(q)
&= -\zeta N \int_0^\infty \id\tau\,
\left\langle
F(r_j-z(\tau)) \, ; \, 
\frac{\partial \log \rho_q}{\partial q_\ell}(z, \sigma)
\right\rangle_q^{\text{BO}}
\label{def friction coefficient} \\
B_{j\ell}(q)
&=\zeta^2N \int_0^\infty \id\tau\,
\left\langle
F(r_j-z(\tau)) \, ; \, 
F(r_\ell-z(0))
\right\rangle_q^{\text{BO}} \label{def B full}
\end{align}
where the single particle Born-Oppenheimer distribution is used, \color{black} such that we can drop the vector notation over $z,\sigma$ \color{black}.
Noting that the covariance function $C_{j \ell}(\tau; q) = \left \langle F(r_j-z(\tau)) \ ; \ F(r_\ell-z(0)) \right\rangle_q^{\text{BO}}$ is even in $\tau$ due to stationarity, it follows that the noise amplitude is always positive since
\begin{align*}
    B_{j \ell}(q) = \zeta^2 N \int_0^{\infty} \id \tau \ C_{j \ell}(\tau; q) = \frac{\zeta^2 N}{2} \int_{- \infty}^{\infty} \id \tau \ C_{j \ell}(\tau;q) = \frac{\lambda^2 N}{2} S_{j \ell}(0;q)
\end{align*}
where $S_{j \ell}(\omega;q) =\int_{- \infty}^{\infty} \id \tau \ C_{j \ell}(\tau;q) \ e^{i \omega \tau } $ is the spectral density and we used $S_{j \ell}(\omega;q) \geq 0$ (Wiener–Khinchin theorem \cite{kamp}). \\

In the equilibrium limit, $v_0 \to \infty , \ \alpha \to \infty$ with $v_0^2/\alpha = 2\mu_z k_B T$ constant, the Born-Oppenheimer distribution has the Boltzmann form $\rho_q(z) = \sum_{\sigma} \rho_q(z,\sigma) = e^{- \beta \zeta  U_{\text{int}}(q,z)}/Z_q$, yielding the equilibrium second fluctuation--dissipation theorem (FDRII), \cite{balakrishnan2020elements,maes2014second}
\begin{align}
     \nu_{j \ell}(q) & = \beta B_{j\ell}(q) \label{fdr2}
\end{align}
  In particular, the friction is always a positive quantity since $ B_{j\ell}$ is. The main question of the present paper is to understand when and how all that changes when out of equilibrium. \\ 
In the general case, the averages Eqs.~\eqref{mean force}--\eqref{def B full} cannot be computed exactly, and we restrict our analysis to the weak coupling regime $\zeta \ll 1$ up to $O(\zeta^2)$.

\subsection{Streaming term}\label{subsection streaming term}
As computed in Appendix \ref{appendix streaming term}, the streaming term Eq.~\eqref{mean force} becomes up to order $O(\zeta^2)$
\begin{align}
\zeta \bar{F}(q)
&=
\zeta \frac{N}{L} \oint  \id x \ F(x)
-\zeta^2 N \beta_{\text{eff}}
\sum_{\ell=0}^{n-1}{\cal M}_{r_j-r_\ell}q_\ell
+O(\zeta^3)
\label{streaming weak coupling} \\
{\cal M}_r &= \frac{1}{L} \oint \id x \  F(r+x) \ F(x) -\left(\frac{1}{L} \oint \id x \  F(x) \right)^2 \nonumber 
\end{align}
For the von Mises form Eq.~\eqref{von mises}, one finds explicitly
\begin{align*}
    &\frac{1}{L} \oint  \id x \ F(x) = F_0, \qquad \cal M_r 
    = F_0^2 \left(\frac{I_0\left(2 p \cos\left(\frac{\pi r}{L} \right) \right)}{I_0(p)^2}  - 1 \right)
\end{align*}
and
\begin{align*}
    \lim_{p \to 0} \cal M_r = 0, \qquad \lim_{p \to \infty} \cal M_r = F_0^2 \left(\sum_{k \in \mathbb{Z}} L \ \delta(r-kL) -1\right)
\end{align*}
Only focusing on this term, the reduced dynamics Eq.~\eqref{reduced dynamics 1} becomes
\begin{align*}
     M\frac{\id^2 q_{j}}{\id t^2} &= \kappa (q_{j+1} + q_{j-1} - 2 q_{j}) - \sum_{\ell=0}^{n-1} \left(K \delta_{j \ell} - \zeta^2 N \beta_{\text{eff}} {\cal M}_{r_j-r_\ell}\right) \  q_\ell  - \zeta \frac{N}{L} \oint  \id x \ F(x)
\end{align*}
\color{black} Here, \color{black} $\zeta \frac{N}{L} \oint  \id x \ F(x) = \zeta N F_0$ \color{black} represents \color{black} a constant force \color{black} resulting in an overall static deflection 
\begin{align}\label{static deflection}
q^* = - \frac{\zeta N F_0}{K}    
\end{align}
of the chain away from the origin; see Appendix \ref{appendix streaming term} for a derivation. This effect is visible \textit{e.g.} in the phase space trajectories in Fig. \ref{mean phase space trajectories}. \\
Next, \color{black} the term $\zeta^2 N \beta_{\text{eff}} {\cal M}_{r_j-r_\ell}$ renormalizes the harmonic confinement (Lamb shift \cite{openquantumtheory}). \color{black}The linearity in $q$ of the $O(\zeta^2)$ term appears because (i) the average in the mean force is computed perturbatively in the coupling parameter $\zeta \ll 1$, and (ii) the interaction potential between the chain and the bath in Eq. \eqref{Uint} is linear in $q$, rather than a linearization procedure for small $q$.  More generally, if the coupling potential were proportional to $q^s$, then the $O(\zeta^2)$ term in the mean force would include $q^s$ as well.  \color{black} \\
This \color{black} reduced dynamics \color{black} can also be interpreted in Fourier series using the discrete Fourier transform, \cite{dftbookt},
\begin{align}\label{dft transform}
q_j(t)=\sum_{k=0}^{n-1}\tilde q_k(t)e^{2\pi i k j/n},
\qquad
\tilde q_k(t)=\frac{1}{n}\sum_{j=0}^{n-1}q_j(t)e^{-2\pi i k j/n}
\end{align}
resulting in (for $k \in \{0,...,n-1\}$)
\begin{align}
M\frac{\id^2\tilde q_k}{\id t^2}
&=
-M\omega_k^2\tilde q_k
-\zeta N \delta_{k,0}\tilde F_0 \label{fourier space mean force only}
\\
M\omega_k^2
&=
2\kappa\left(1-\cos\frac{2\pi k}{n}\right)
+K_{\mathrm{eff},k}, \qquad 
K_{\mathrm{eff},k} = K- \left(1-\delta_{k,0}\right)\zeta^2N n \beta_{\text{eff}}
\sum_{\ell\in\mathbb Z}
|\tilde F_{k+\ell n}|^2   \label{omegak}
\end{align}
with $\tilde{F}_k = \frac{1}{L} \oint \id x \  F(x) \ e^{- 2 \pi i k x/L}$ the $k$th Fourier coefficient of $F$. As such, in Fourier space, the modes are uncoupled \color{black} harmonic oscillators, \color{black} which makes it convenient to look at them (only). The active bath reduces the effective confinement $K_{\text{eff},k} < K$ of all non-zero modes, while only the zeroth mode \color{black}(COM) \color{black} receives the constant force proportional to $\tilde F_0$, \color{black} shifting the equilibrium position of the oscillator by $-\frac{\zeta N \tilde F_0}{K}$. \color{black} 

\subsection{Friction and noise}
The noise amplitude and friction are computed in Appendices \ref{appendix noise term}, \ref{appendix friction term} with the result up to order $O(\zeta^2)$
\begin{align}\label{friction and noise}
\nu_{j\ell}&=\sum_{a=1}^{\infty}\cal V_a
\cos\left[\frac{2\pi a}{n}(j-\ell)\right],
\qquad
B_{j\ell}=\sum_{a=1}^{\infty}  \cal B_a
\cos\left[\frac{2\pi a}{n}(j-\ell)\right] \\
 \cal V_a &= \frac{2\zeta^2N\mu_z}{v_0^2} |\tilde F_a|^2 \left( \frac{\alpha^2L^2} {\pi^2v_0^2a^2}-1 \right) + O(\zeta^3) , \qquad \cal B_a = \zeta^2 \frac{N L^2\alpha}{\pi^2 v_0^2} \frac{|\tilde F_a|^2}{a^2} + O(\zeta^3) \label{V a B a}
\end{align}
\begin{align*}
    \nu_{j\ell}&= \frac{2\zeta^2N\mu_z}{v_0^2} \sum_{a=1}^{\infty}  |\tilde F_a|^2 \left( \frac{4\alpha^2} {v_0^2 k_a^2}-1 \right)
\cos\left[\frac{2\pi a}{n}(j-\ell)\right]
\end{align*}
 Hence, the noise covariance is positive definite in the sense that
\begin{align*}
    \frac{1}{n^2} \sum_{j, \ell = 0}^{n-1} x_{j } \ B_{j \ell } \ x_{\ell } &= \zeta^2 \frac{N L^2 \alpha}{\pi^2 v_0^2} \sum_{a = 1}^{\infty} \frac{|\tilde{F}_a|^2}{a^2} |\tilde{x}_a|^2  \geq 0
\end{align*}
 while the friction satisfies
\begin{align*}
    \frac{1}{n^2} \sum_{j, \ell = 0}^{n-1} x_{j} \ \nu_{j \ell} \ x_{\ell } & = \frac{2\zeta^2 N\mu_z}{v_0^2} \sum_{a = 1}^{\infty}  |\tilde{F}_a|^2 \ |\tilde{x}_a|^2  \left(\frac{\alpha^2 L^2}{ \pi^2 v_0^2 a^2} -1\right)% = \frac{2\zeta^2 N\mu_z}{v_0^2} \sum_{a = 1}^{\infty}  |\tilde{F}_a|^2 \ |\tilde{x}_a|^2  \left(\frac{4 \alpha^2}{ k_a^2 v_0^2} -1\right)
\end{align*}
which does not have a fixed sign. 
%\textcolor{red}{May be we clarify here what $a$ is? some wavevector $k_a=2\pi a/L$, then shorter wavelength modes $<2\alpha/v_0$ can still have negative friction, the criteria that we give is for all modes to be negative. Moreover, intuitively also, if $L\gg v_0/\alpha$ the dynamics of the bath particles almost become passive like} 
Yet, $\nu$ is negative-definite for 
\begin{align}\label{critical alpha}
    \alpha < \alpha_c = \frac{\pi v_0}{L} 
\end{align}
which is the same criterion as in \cite{beyen2025couplingelasticstringactive}. That implies a linear instability in the reduced dynamics when the RTPs are far from equilibrium, \textit{i.e.} they are very persistent. \color{black} This criterion represents a competition between the persistence length of the RTPs, $L_{\text{RTP}} = v_0/(2\alpha)$, and the length $L$ of the ring, \textit{i.e.} $L_{\text{RTP}} $ must be larger than $L/(2 \pi)$, the inverse of the smallest wave number $2 \pi/L$. From the point of view of a long chain, $L \gg L_{\text{RTP}}$, the bath behaves essentially passively, and no negative friction can be induced.  
\color{black} \\
\color{black} In more detail, as derived in Appendix \ref{appendix langevin equation}, when going to Fourier space, the full reduced dynamics in Fourier space becomes
\begin{align}
    & M\frac{\id^2 \tilde{q}_k}{\id t^2} + \tilde \nu_k\frac{\id \tilde{q}_k}{\id t} + M\omega_{k}^2 \tilde{q}_k = -\zeta \delta_{k,0} N \tilde{F}_0 + \tilde{\xi}_{k} \\
    & \left \langle \tilde{\xi}_{k} \right \rangle = 0, \qquad \left \langle \tilde{\xi}_{k}(t) \ \tilde{\xi}_{m}(t') \right \rangle =  2 \tilde{B}_{k} \ \delta_{k,m} \delta(t-t') \nonumber \\ 
    &\tilde{\nu}_k= \frac{n}{2}(1-\delta_{k,0})\sum_{a \in \mathbb{Z}} \cal V_{k + a n}, \qquad \tilde{B}_{k} = \frac{(1 -\delta_{k,0}) }{2} \sum_{a \in \mathbb Z} \cal B_{k + a n} \nonumber 
\end{align}
with $\omega_k$ from Eq.~\eqref{omegak} and $\cal V_a, \cal B_a$ from Eq.~\eqref{V a  B a}. Consequently, each mode behaves independently as an (anti)-damped harmonic oscillator. In particular, $ \tilde{\nu}_k$ represents the friction coefficient experienced by the $k$th mode and $\tilde \nu_0 = 0$, \textit{i.e.} the COM does not experience any friction. Hence, for given $\alpha, L, v_0$ all modes $k$ with
\begin{align}\label{mode criterion}
  |k| > \frac{\alpha L}{ \pi v_0} % k^2 \geq \frac{\alpha^2 L^2}{ \pi^2 v_0^2}
\end{align}
experience negative friction since then $\tilde{\nu}_k < 0$, while other modes have $\tilde{\nu}_k > 0$. Hence, large modes are always negatively damped, while only lower modes can experience positive friction. The condition \eqref{critical alpha} then corresponds to the case where all modes (except $k = 0$) experience negative friction.
\color{black} \\

\color{black} Comparing with the equilibrium FDRII result \eqref{fdr2}, the friction \eqref{friction and noise} naturally splits into a part proportional to $B_{ij}$, but also a new, nonequilibrium correction which we denote as $\gamma_{ij}$, \textit{i.e.}, \color{black}
%It is interesting to write the friction in \eqref{friction and noise} into the form
\begin{align}
\nu_{\color{black} ij \color{black}} = \beta_{\mathrm{eff}} B_{\color{black} ij \color{black}}+\gamma_{\color{black} ij \color{black}} + O(\zeta^3) , \qquad 
\gamma_{j\ell} =\sum_{a=1}^{\infty}\cal G_a \cos\left[\frac{2\pi a}{n}(j-\ell)\right],
\qquad \cal G_a = -\frac{2\zeta^2N\mu_z}{v_0^2}|\tilde F_a|^2 \label{entropic frenetic split}
\end{align}
\color{black} Here, $\beta_{\mathrm{eff}}=\frac{2\alpha\mu_z}{v_0^2}$ is the well-known effective temperature for a RTP \cite{Santra_2023}, \color{black}
%the first term is the entropic part proportional to the noise amplitude \color{black} $B_{ij}$ in \eqref{friction and noise} \color{black} at effective temperature $\beta_{\mathrm{eff}}=\frac{2\alpha\mu_z}{v_0^2}$ as in FDRII, 
whereas $\gamma$ is \color{black} the active contribution 
%the nonequilibrium correction 
generated by \color{black} the \color{black} persistence of the active particles \color{black} which \color{black} is always negative definite, \color{black} breaking the equilibrium FDRII. \color{black} It is precisely this active \color{black} correction \color{black} that can make the total friction negative, leading to a linear instability. \color{black} There are of course many other ways of splitting the friction in two parts, but Eq. \eqref{entropic frenetic split}  has the particularly nice feature of splitting it into the entropic friction $\nu_{ij}^{\text{ent}} = \beta_{\mathrm{eff}} B_{\color{black} ij \color{black}}$ and frenetic part $\nu_{ij}^{\text{fren}} = \gamma_{ij},$ which is insightful for nonequilibrium considerations; see \cite{pei2024inducedfrictionprobemoving, beyen2025couplingelasticstringactive, frenesy, maesresponse} for more details.  \color{black} \\

\begin{figure}[H]
    \centering
    \includegraphics[width=0.95\linewidth]{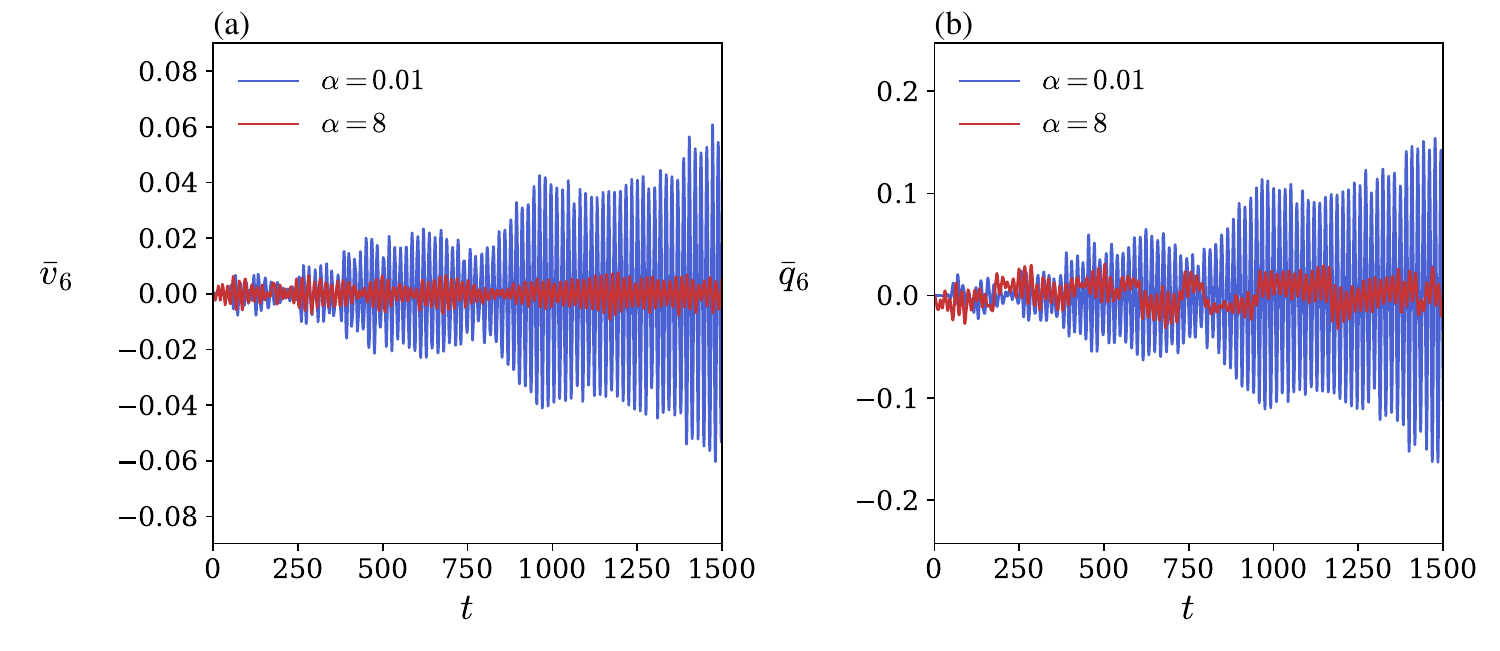}
    \caption{Time evolution of the $k=6$ Fourier mode in the transient regime. \color{black}(a): \color{black} velocity mode $\tilde v_6(t)=\dot{\tilde q}_6(t)$; \color{black} (b): \color{black} displacement mode $\tilde q_6(t)$. For all the plots, we have chosen $n=10$, $N=10$, $L=10$, $v_0=4$, $\kappa=0.1$, $K=1$, \color{black} $\zeta=0.01$, $p=2$, $F_0=10$ \color{black} (arbitrary units) \color{black} such that the criterion \eqref{mode criterion} yields an instability for modes $|k| > 0.796 \alpha$ (arbitrary units). For $k = 6$, this inequality implies negative friction and hence a growing mode for $\alpha < 7.540 $ (arbitrary units) and positive friction otherwise, in agreement with the simulation results. \color{black}}
    %The growth of the mode amplitude at small $\alpha$ illustrates the negative-friction instability predicted by the reduced Langevin dynamics.}
    \label{f:modes}
\end{figure}

Fig.~\ref{f:modes} shows the time evolution of a selected mode in regimes of positive and negative induced friction. In the negative-friction regime, the mode amplitude grows, confirming the onset of the linear instability. As the oscillator amplitudes and velocities increase, the assumptions underlying the reduced dynamics eventually break down; in particular, the time-scale separation between the slow chain and the fast bath particles ceases to be true. This raises the central question for the remainder of the paper: does this instability lead to an unbounded growth of the chain, or is it arrested by nonlinear effects beyond the reduced linear theory? We address this in the following via numerical simulations. 

\section{Saturation regime}\label{section saturation regime}

\subsection{Long time trajectories}
\begin{figure}[H]
    \centering
    \includegraphics[width=\linewidth]{
    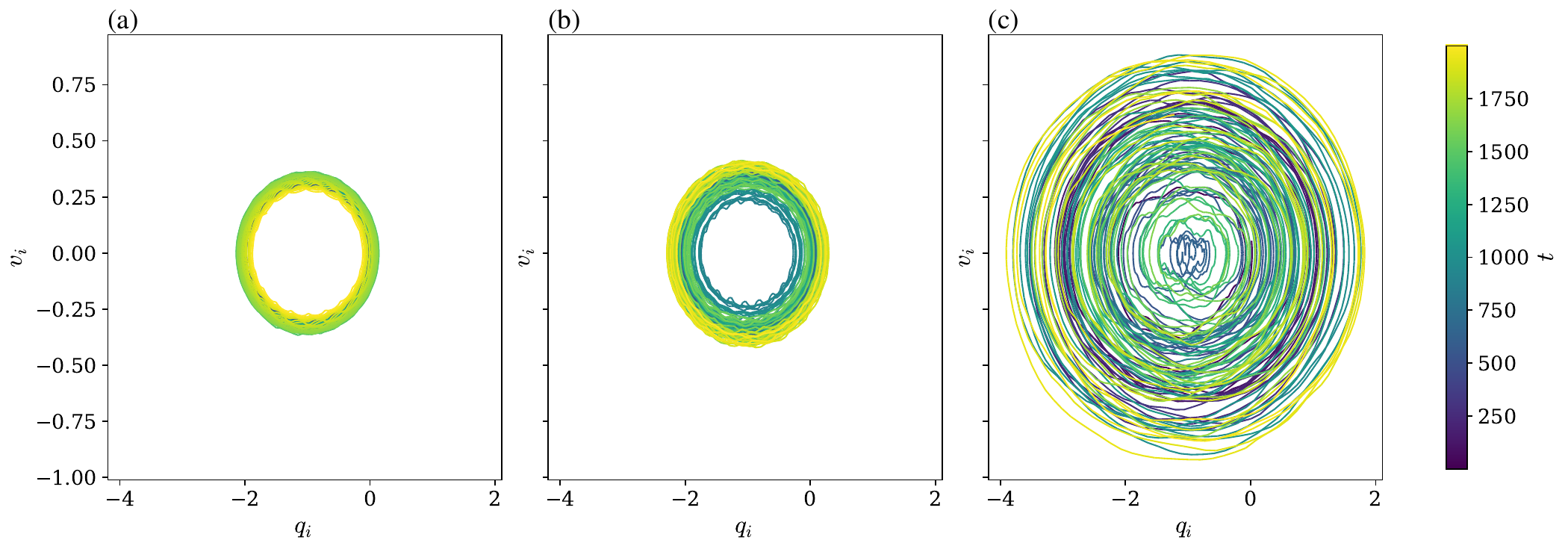}
    \caption{\raggedright Phase-space trajectories of a tagged oscillator for different values of the bath activity. From \color{black}(a) to (c)\color{black}, $\alpha=0.01$, $\alpha=0.1$, and $\alpha=8$. For all panels, we have chosen the parameters: $n=10$, $N=10$, $L=10$, $v_0=10$, $M=10$ $\kappa=0.1$, $K=1$, \color{black} $\zeta=0.01$, $p=2$, $F_0=10$ \color{black} and initial conditions $q_i(0)=0$, $v_i(0)=0.1$ (arbitrary units)  \color{black} such that the critical value in \eqref{critical alpha} yields $\alpha_c = \pi$ and $q^* = -1$ (arbitrary units). For the simulations in (a) and (b), we have $\alpha < \alpha_c$ with the corresponding negative friction effects, while simulation (c) at $\alpha > \alpha_c$ does not have these properties. We also note that the shifted point $(q^\ast,0)$ from \eqref{static deflection} exactly corresponds to the center of the circles. \color{black}}
    \label{mean phase space trajectories}
\end{figure}

To understand the fate of the oscillator chain at large times, we first look at the phase space trajectories of a tagged oscillator. This is illustrated in Fig.~\ref{mean phase space trajectories} for different values of the bath activity \color{black} $\alpha$. As already explained in section \ref{subsection streaming term}, the limit cycles are not centered around the origin but shifted by the amount \eqref{static deflection} towards negative $q$ due to the constant force $\frac{\zeta N}{L} \oint \id x \ F(x)$ in Eq.~\eqref{streaming weak coupling}. \color{black} 
For small $\alpha$, the long-time phase-space trajectories of the tagged oscillator \color{black} are depleted around $(q^\ast,0)$ and concentrate along an orbit of a finite radius, \color{black} \textit{i.e.} they form an annulus, \color{black} reminiscent of a limit cycle. \color{black} In contrast, for large $\alpha$, i.e., weak activity/persistence, the trajectories do not enclose a clear structure and remain localized near \color{black} $(q^*,0)$, \color{black} consistent with positively damped oscillatory motion with fluctuations. \\
The closed orbits in the phase-space trajectories seen here are reminiscent of Rayleigh oscillators described by the dynamics, \cite{Rayleigh01041883,Chen_1994, quantumrayleigh},
\begin{equation}\label{Rayleigh oscillator}
m \ddot x(t) + k x(t) =\gamma_1 \dot x(t) - \gamma_2 \dot x(t)^3, \qquad \gamma_1, \gamma_2 > 0  
\end{equation}
where initial small velocities are amplified due to a linear negative friction, while they become damped at higher velocities due to positive non-linear friction. This model is consistent with the transient negative friction regime for $\alpha<\alpha_c$. However, at larger values of $\alpha$, the induced friction is positive in linear order, and no \emph{active} signatures of the bath are seen anymore. \color{black} One also observes in Fig. 3 that the ring-like structure is more concentrated, and hence has a smaller radius, for smaller $\alpha$ (larger activity), while at larger $\alpha$, the trajectories spread out more, increasing the effective radius.
%. However, although a Gaussian has more spread, it does spend more time in the central region of the circle, while for small $\alpha$, the particle is more concentrated at the boundaries. This observation explains why the saturated value of the MSD in the next subsection will be higher for smaller $\alpha$ than the Gaussian case, so that the radius and MSD don't necessarily agree with each other. \color{black}
Having established the existence of a long-time nonequilibrium stationary state (NESS), we focus on more details of this stationary regime.
\subsection{Velocity and displacement fluctuations of a tagged oscillator}
\begin{figure}[H]
    \includegraphics[width=0.47\linewidth]{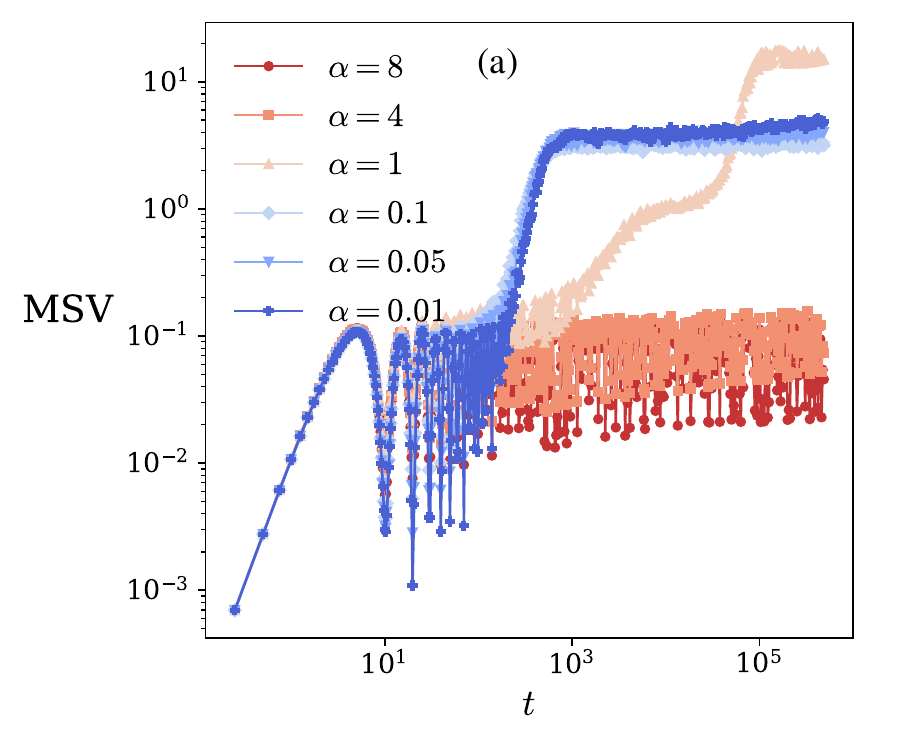}
    \includegraphics[width=0.47\linewidth]{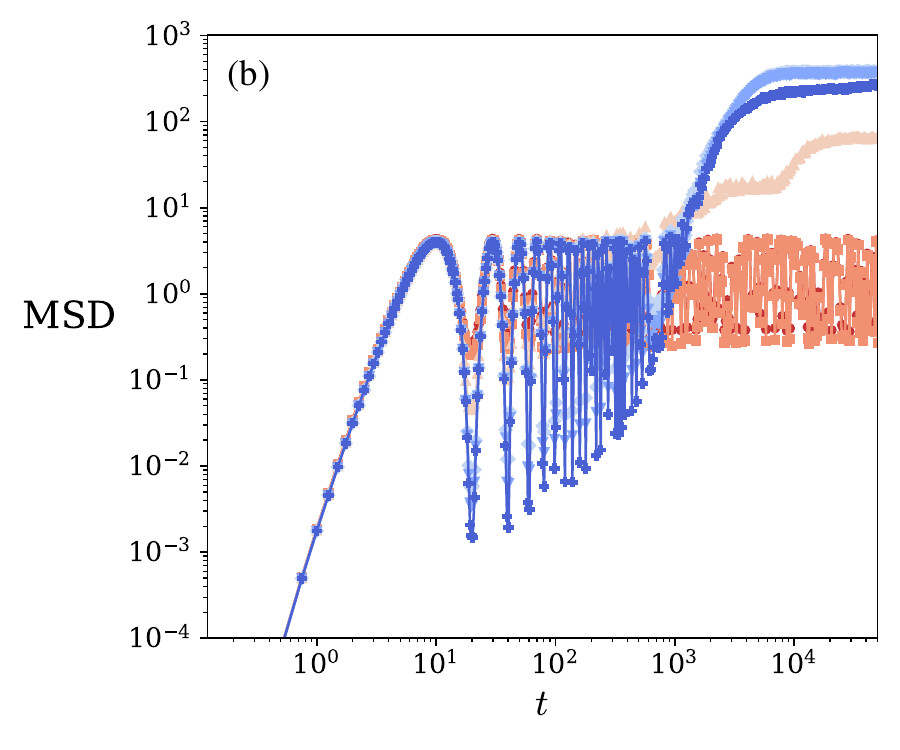}
  \caption{\raggedright  Temporal evolution of the mean-squared velocity \color{black} (a) \color{black} and mean-squared displacement \color{black} (b) \color{black} from numerical simulations. Here we considered 10 oscillators with $M = 10, \kappa = 0.1$ $K=1$, \color{black} $\zeta=0.01$, $p=2$, $F_0=10$ \color{black} and $N = 10$ bath particles with $v_0=2,L=25$ (arbitrary units).}
  \label{fig:msdmsv}
\end{figure}

A first way to understand the approach to\color{black}, \color{black} and properties of\color{black}, \color{black} the NESS is through the mean-squared velocity (MSV) and the mean-squared displacement (MSD) of a tagged oscillator. For the $i$th oscillator with velocity $v_i(t)=\dot q_i(t)$, these observables are defined by,
\begin{equation}\label{msd msv}
\text{MSV}=\left\la [v_i(t)-v_i(0)]^2 \right\ra, \qquad  \text{MSD}=\left\la [q_i(t)-q_i(0)]^2 \right\ra
\end{equation}
where the brackets $\left \langle \cdot \right \rangle$ without subscript indicate averages over the appropriate noise, here over the $\sigma$-realizations. We also got rid of the index $i$ on the left-hand side since we are on a ring, and there is translational invariance. That also follows from the reduced dynamics since all functions depend only on the distance $(j-\ell)/n = (r_j-r_\ell)/L$ between sites on the ring. \\
The observables in Eq.~\eqref{msd msv} are standard measures of fluctuations that are also directly accessible in experiments. If the negative-friction regime produced a true runaway solution, both quantities would continue to grow without bound in time, while if there is a stationary state, both will approach finite plateaus. The numerical results are shown in Fig.~\ref{fig:msdmsv} and one observes that both the MSV and the MSD eventually saturate for all considered values of $\alpha$, \textit{i.e.} the transient instability eventually stops and crosses over to a stationary nonequilibrium regime at long times.
At the same time, the relaxation towards the stationary state is remarkably slow compared to the microscopic time scales $\alpha^{-1}$, $\omega_{\mathrm{chain}}^{-1}$, and $L/v_0$, and exhibits a multi-step relaxation before reaching its final saturation value. The first growth regime is essentially independent of $\alpha$ and reflects the short-time inertial response of the oscillator inside the harmonic trap, namely, $\mathrm{MSV}\sim t^2$,$\mathrm{MSD}\sim t^4$. \color{black} A mode $k \neq 0$ behaves as a (anti-)damped harmonic oscillator with noise
and therefore oscillates with period $T_k=2\pi/\omega_k=O(\zeta^0)$, while its
amplitude envelope evolves as $ \propto e^{-\tilde\nu_k t/2M}$ due to the induced frictional force $-\tilde \nu_k \dot{\tilde q}_k$, , i.e. on
\begin{equation*}
\tau_k = \frac{2M}{\tilde\nu_k} = O(\zeta^{-2}).    
\end{equation*}
%gives rise to a characteristic velocity decay/growth $v(t) \propto e^{-\nu t/m}$, defining the frictional timescale $\tau_{\text{friction}} = m/\nu$. 
Since $\tilde \nu_k = O(\zeta^2)$ is small in the weak coupling regime $\zeta \ll 1$, this timescale is large, so long observation times are needed for friction to have an appreciable effect.
\color{black}
%The induced frictional force \color{black} gives rise to a characteristic velocity decay/growth $v(t) \propto e^{-\nu t/m}$, defining the frictional timescale $\tau_{\text{friction}} = m/\nu$. Since $\nu = O(\zeta^2)$ is small, the friction acts \color{black}
%is an $O(\zeta^2)$ effect and therefore acts only 
%on a much longer time scale in the weak coupling regime. \color{black}  explain more \color{black} 
As a result, the system first reaches an intermediate plateau before the induced friction becomes visible. Beyond the intermediate regime, we observe a second growth stage for sufficiently small $\alpha$, in agreement with the appearance of negative induced friction. This second, much steeper growth is absent for larger values of $\alpha$, where the induced friction remains positive, and the first plateau already corresponds to the final stationary value. The final saturation at small $\alpha$ is then due to nonlinear correction terms that become effective when the displacements and their velocities have grown sufficiently, akin to the Rayleigh oscillator Eq.~\eqref{Rayleigh oscillator}. This provides a simple explanation for why the relaxation times are so large. \\ 
It is also interesting to note that the stationary values of the MSV and MSD are non-monotonic functions of $\alpha$. For example, the final MSV and MSD values of $\alpha =0.05$ are larger than those of $\alpha = 0.1$ and $\alpha = 0.01$. The largest long-time fluctuations are found in the regime where the persistence length $v_0/\alpha$ becomes comparable to the system size $L$, {\it i.e.}, when
 $\frac{v_0}{\alpha L}\sim O(1)$. 
For larger tumbling rates, $(\alpha \gtrsim v_0/L)$, the stationary values decrease monotonically with an increase in $\alpha$. The strongest enhancement of fluctuations is therefore not found deep in the short-persistence regime, but near the crossover where persistence and finite-size effects compete most efficiently. 

\subsection{Stationary distributions}
\begin{figure}[ht]
    \includegraphics[width=0.47\linewidth]{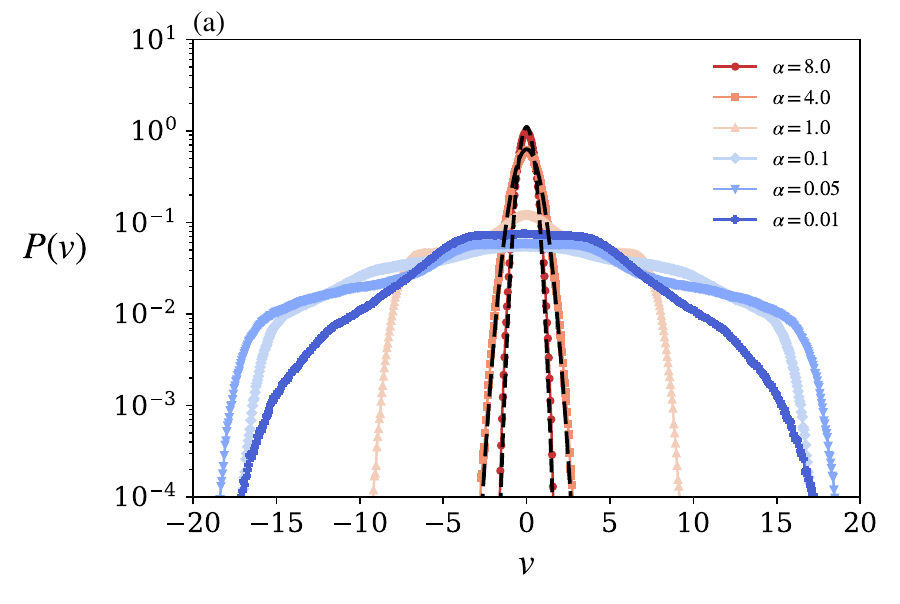}
    \includegraphics[width=0.47\linewidth]{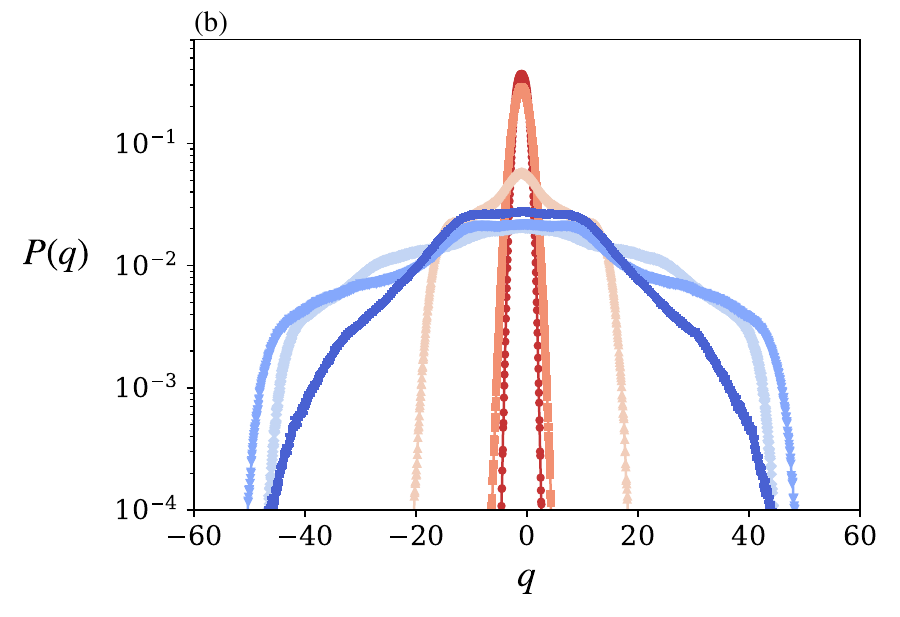}
\caption{\raggedright Stationary distributions of a tagged oscillator. \color{black} (a) \color{black}: velocity distribution $P(v_i)$. \color{black} (b) \color{black}: displacement distribution $P(q_i)$. For both panels, we have chosen the parameters: $n=10$, $N=10$, $M=10$, $\kappa=0.1$, $K=0.2$, $v_0=4$, $L=25$, \color{black} $\zeta=0.01$, $p=2$, $F_0=10$ \color{black}. \color{black} The dashed lines correspond to the Gaussian equilibrium distribution \eqref{gaussian eff temperature} at temperature $k_B T_{\text{eff}} \to k_B T$. \color{black}}
  \label{fig:dists}
\end{figure}

The second moments MSV and MSD discussed above provide a first characterization of the stationary regime. Here we add a discussion of the stationary one-point distributions of the position and velocity of a tagged oscillator, shown in Fig.~\ref{fig:dists}. As shown there, for sufficiently large $\alpha$, both the velocity and the position distributions are close to Gaussian (as for the equilibrium harmonic chain). This behaviour can be explained as follows. For $t \gg \alpha^{-1}$, the noise autocorrelation function Eq.~\eqref{auto correlation} reduces to, \cite{Basu_Majumdar_Rosso_Schehr_2019,santra2022universal},
\begin{align*}
    \left \langle v_0 \sigma(t) \ v_0 \sigma(t') \right \rangle = v_0^2 e^{- 2 \alpha |t-t'|} \to 2 D_{\text{eff}}\, \delta(t-t') 
\end{align*}
with an effective diffusion constant $D_{\text{eff}} = v_0^2/(2 \alpha)$. Hence, for $t \gg \alpha^{-1}$, the active particle effectively reduces to a passive particle. In the regime $\alpha \gg 1$, this happens for almost all times, and the tagged oscillator behaves approximately as if it were coupled to an effective thermal reservoir, \textit{i.e.}, a velocity distribution of the form
\begin{align}\label{gaussian eff temperature}
    P(v_i)= \sqrt{\frac{M}{2\pi k_BT_{\text{eff}}}}e^{-\frac{Mv_i^2}{2k_BT_{\text{eff}}}} \quad\text{ with   } \quad  k_BT_{\text{eff}}=v_0^2/(2\alpha \mu_z)
\end{align}
Thus $\alpha^{-1}$ plays the role of an “activity” parameter—as $\alpha$ increases from zero, the process crosses over from a strongly active regime ($\alpha \to 0$) to a strongly passive one ($\alpha \to \infty$). In the active regime, the distributions in Fig. \ref{fig:dists} exhibit broad shoulders which depend on  $\alpha$ in a non-monotonic way \textit{i.e.} they are the widest in the same parameter range where the stationary MSV and MSD are largest. These shoulders are the distribution-level signature of the enhanced stationary fluctuations already seen in Fig.~\ref{fig:msdmsv}. Moreover, they indicate that the tagged oscillator explores a wider region in phase space (\textit{e.g.} reaches larger velocities) than in equilibrium, which is another indication of the negative linear friction regime. For still larger $\alpha$, these shoulders gradually disappear, and the distributions become narrower, eventually converging to the expected Gaussian form Eq.~\eqref{gaussian eff temperature}. The takeaway is that the stationary state %is stationary, but its 
fluctuations retain signatures of the linear instability giving rise to the broad non-Gaussian stationary distributions. 

\subsection{Stationary spatial correlations}

The one-point distributions discussed above characterize the size of the stationary fluctuations for a fixed oscillator with index $i$, but they do not tell us whether neighbouring oscillators move independently or in a coordinated way. To probe the collective structure of the long-time state, we also include the equal-time spatial correlations of the velocity and displacement fields. These correlations are important because they show whether the saturation regime is only locally fluctuating, or whether it also develops coherent spatial organization along the chain.

\color{black} Given the observables $O_i = q_i, v_i$, \color{black} we define the normalized steady-state correlations \color{black} $C_O(i, \ell)$ \color{black} as
\begin{equation}
C_v(\color{black}i,\color{black}\ell )=\frac{\la v_i \, ; \, v_{i+\ell}\ra}{\la v_i \, ; \, v_i\ra},
\qquad
C_q(\color{black}i,\color{black}\ell)=\frac{\la q_i \, ; \, q_{i+\ell}\ra}{\la q_i \, ; \, q_i\ra}
\label{eq:cv-cq-def}
\end{equation}
where $\la O_i \, ; \, O_{i+\ell }\ra=\la O_iO_{i+\ell}\ra-\la O_i\ra\la O_{i+\ell}\ra  $, and the averages are taken in the stationary state. The normalization ensures
$C_{\color{black} O \color{black}}(\color{black}i,\color{black}0)=1$,
so that the spatial structure can be easily compared directly for different values of the persistence rate. Due to the periodicity of the chain, we also have $C_{\color{black} O \color{black}}(\color{black}i,\color{black}\ell+n) = C_{\color{black} O \color{black}}(\color{black}i,\color{black}\color{black} \ell \color{black})$ for $n$ oscillators. \color{black} Moreover, since the chain is homogeneous in the steady state, the correlations $C_O = C_O(\ell)$ do in fact not depend on $i$, so relabelling $i \to i-\ell$ gives $C_O(\ell) = \la O_{i-\ell} \, ; \, O_{i}\ra/\la O_i ; O_i \ra = \la O_{i} \, ; \, O_{i-\ell}\ra/\la O_i ; O_i \ra = C_O(-\ell)$, \textit{i.e.} the equal-time autocorrelations are even in $\ell$. Combined with the periodicity, $C_O(n-\ell) = C_O(-\ell) = C_O(\ell)$, so $C_{q,v}$ is mirror-symmetric about the point of maximum separation $\ell = n/2$ and, \color{black} it suffices to \color{black} discuss and \color{black} plot $C_{q,\color{black} v \color{black}}(\ell)$ for $\ell \in \color{black} [0,n/2] \color{black}$. \\

We first recall that for an uncoupled chain in equilibrium, the velocity distribution of the harmonic chain has the Maxwell-Boltzmann form, which is diagonal in the site velocities, \textit{i.e.} the equal-time velocity correlation is strictly local, \cite{landau1980statistical}
\begin{equation}
\la v_i \, ; \,  v_{i+\ell}\ra_{\mathrm{eq}}=\frac{k_B T}{M}\,\delta_{\ell,0}
\label{eq:eq-vel-corr-final}
\end{equation}
which implies $C_v^{\mathrm{eq}}(\ell)=\delta _{\ell,0}$.
Indeed, in equilibrium, there are no equal-time velocity correlations between different oscillators. Moreover, in equilibrium, the position distribution has the Boltzmann form $\rho_{\text{eq}}(q) \propto e^{- \beta E(q)}$ with energy from Eq.~\eqref{E(q)}. By going to Fourier space with Eq.~\eqref{dft transform}, one finds 
\begin{align*}
    E(q) =\frac{n}{2} \sum_{k = 0}^{n-1} \left[K + 2 \kappa \left(1 - \cos\left(\frac{2 \pi k}{n}\right) \right) \right] |\tilde{q}_k|^2
\end{align*}
such that for $n$ sufficiently large (as we deal with an extended chain), one finds the real-space autocorrelation,
\begin{equation}
\la q_i \,; \, q_{i+\ell}\ra_{\mathrm{eq}}
=
\frac{k_BT}{2\pi}\int_{-\pi}^{\pi}\frac{e^{ik \ell}\,dk}{K+2\kappa(1-\cos k)}
\label{eq:eq-pos-corr-final}
\end{equation}
For a pinned chain with $K>0$, the large-distance behavior ($k\to 0$) is a monotonic exponential decay towards \color{black} the point of maximum separation \color{black} $\ell = n/2$,
\begin{equation}
\la q_i ;q_{i+\ell}\ra_{\mathrm{eq}} \sim \begin{cases}
    e^{-\ell/\xi_{\mathrm{eq}}}, 
\qquad\text{ with }\ell\in [0,n/2) \\
e^{-(n-\ell)/\xi_{\mathrm{eq}}}, 
\qquad\text{ with }\ell\in [n/2,n)
\end{cases} \qquad 
\xi_{\mathrm{eq}}=\sqrt{\frac{\kappa}{K}}
\label{eq:eq-pos-corr-asym-final}
\end{equation}
and therefore the normalized correlation $C_q^{\mathrm{eq}}(\ell)$ is also monotone. 

\begin{figure}
    \includegraphics[width=0.47\linewidth]{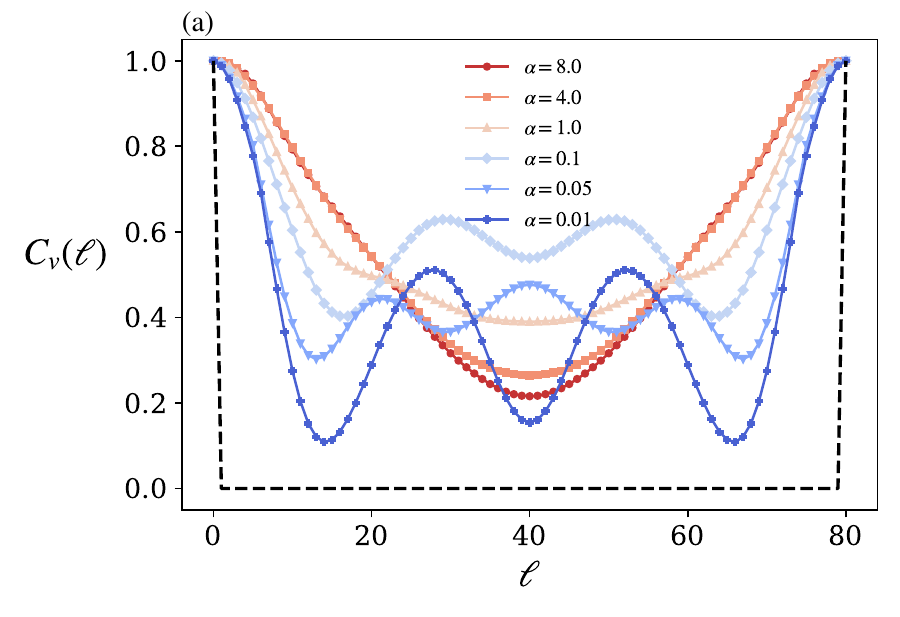}
    \includegraphics[width=0.47\linewidth]{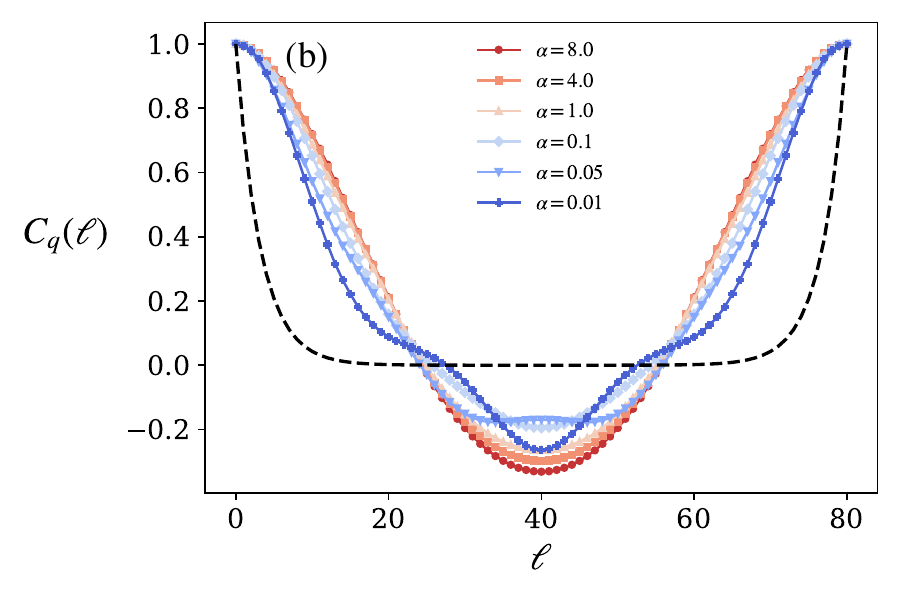}
 \caption{\raggedright Equal-time spatial correlations in the stationary regime. \color{black} (a): \color{black} normalized velocity correlation $C_v(\ell)$, \color{black} (b): \color{black} normalized displacement correlation $C_q(\ell)$ (both defined in Eq.~\eqref{eq:cv-cq-def}). For both panels, we have chosen the parameters: $n=80$, $N=10$, $M=10$, $\kappa=0.1$, $K=0.2$, $v_0=4$, $L=25$, \color{black} $\zeta=0.01$, $p=2$, $F_0=10$ \color{black} (arbitrary units).}
  \label{fig:corr}
\end{figure}

The stationary velocity and position correlations for the oscillator chain are shown in Fig.~\ref{fig:corr}. In contrast to the equilibrium result Eq.~\eqref{eq:eq-vel-corr-final}, the active system always exhibits a finite spatial velocity correlation $C_v(\ell)$. For large values of $\alpha$, it decays exponentially towards $\ell = n/2$, while at smaller $\alpha$, the correlation becomes non-monotonic \color{black} in the region $0 \leq \ell \leq n/2$ (with reflection symmetry around $\ell = n/2$ as discussed above) \color{black} and develops damped oscillations as a function of distance. These oscillations lie within the same parameter regime as the transient negative-friction effect, indicating that the long-time state retains a clear memory of the instability mechanism that drives the early-time growth. Furthermore, the appearance of a non-monotonic decay suggests a damped synchronization along the chain. These structures weaken as $\alpha$ is increased, and disappear completely in the regime where the transient linear instability is absent. \\
The displacement correlations $C_q(\ell)$ show a similar overall trend, with the important difference that in equilibrium, $C_q(\ell)$ already decays exponentially with distance according to Eq.~\eqref{eq:eq-pos-corr-asym-final}. In the active steady state, the decay again becomes non-monotonic in the strongly persistent regime, showing that the positional fluctuations also acquire a collective structure beyond the passive benchmark. However, the oscillatory feature is visibly weaker for $C_q(\ell)$ than for $C_v(\ell)$, which can mainly be attributed to the presence of the on-site harmonic pinning term $K q_i^2/2$, which penalizes large displacements and tends to reduce the spatial propagation of positional order.

\begin{figure}[htb]
    \centering
    \includegraphics[width=0.95\linewidth]{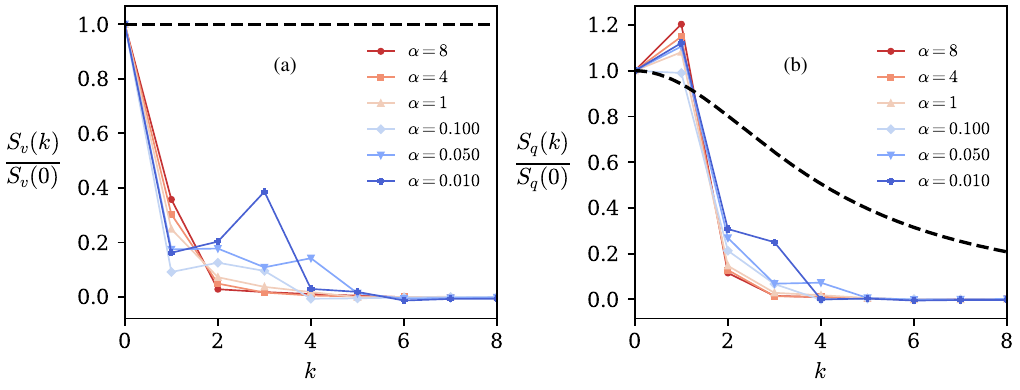}
    \caption{\raggedright Normalized static structure factors. \color{black} (a): \color{black} velocity structure factor $S_v(k)$. \color{black} (b): \color{black} displacement structure factor $S_q(k)$. Both panels are generated by numerical Fourier transform of the same data as in Fig.~\ref{fig:corr}. We show only the first few $k$ modes, where deviations from the equilibrium behavior are most pronounced. \color{black} The dashed lines correspond to the equilibrium structure factor. \color{black}}
    \label{fig:structurefactor}
\end{figure}

A complementary way to characterize the spatial correlations is through the static structure factors, which are accessible in scattering experiments via Fourier analysis of the imaging data. They are obtained from the discrete Fourier transform of the correlations $C_O(\ell)$, where $O\equiv q$ or $v$, \textit{i.e.} 
\begin{align}
    \mathcal{S}_O(k_j)=\sum_{\ell=0}^{n-1}\mathcal{C}_O(\ell)e^{-ik_jr_\ell}=\sum_{\ell=0}^{n-1}\mathcal{C}_O(\ell)\cos(k_jr_\ell)
\end{align}
with wavenumber $k_j=2\pi j/L,~~(j=0,1,2,\cdots, n-1)$. The second \color{black} equality \color{black} follows from the periodic boundary conditions on the chain, \textit{i.e.} $C_O(n-\ell) = C_O(\ell)$ while $\sin(k_j r_{n-\ell}) = - \sin(k_j r_\ell)$. We plot the normalized structure factor $S_O(k)=\mathcal{S}_O(k)/\mathcal{S}_O(0)$ in Fig.~\ref{fig:structurefactor}.
In equilibrium, Eq.~\eqref{eq:eq-vel-corr-final} implies that the velocity structure factor is flat (independent of wavenumber), while the positional structure factor is dominated by long-wavelength modes (see Eq.~\eqref{eq:eq-pos-corr-final}). In contrast, the results in Fig.~\ref{fig:structurefactor} clearly deviate from this. While $S_v(k)$ remains high at small wave numbers, indicating the dominance of long-wavelength collective modes, the velocity structure factor is no longer flat and exhibits a pronounced $k$-dependence. Furthermore, both structure factors display non-monotonic features at finite wave number, which are the Fourier-space counterparts of the oscillatory correlations observed in real space. The non-monotonicities become more prominent for smaller values of $\alpha$, indicating enhanced spatial correlation for higher activity. It is interesting to note that the non-monotonic behavior of the velocity structure factors at finite $k$ has been reported in the context of epithelial monolayers~\cite{henkes2020dense} where peaks or shoulders at finite wave number arise from persistent uncoordinated cell motility coupled to the collective elastic modes of the cell sheets.\\

\begin{figure}[ht]
    \centering
\includegraphics[width=0.48\linewidth]{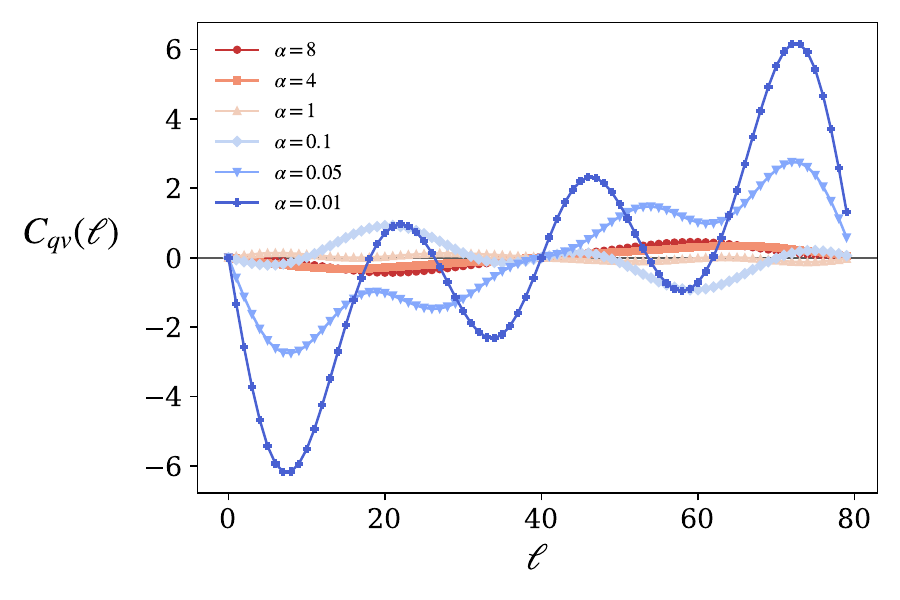}
    \caption{\raggedright Equal-time displacement-velocity cross-correlations $C_{qv}(\ell)$ in the stationary regime for the parameters: $n=80$, $N=10$, $M=10$, $\kappa=0.1$, $K=0.2$, $v_0=4$, $L=25$, \color{black} $\zeta=0.01$, $p=2$, $F_0=10$ \color{black} (arbitrary units).}
    \label{fig:pqcorrelation}
\end{figure}

Finally, we consider the displacement--velocity stationary correlation,
\begin{align}
    C_{qv}(\ell)=\la q_i \, ; \, v_{i+\ell} \ra
    \label{def:crosscorr}
\end{align}
shown in Fig.~\ref{fig:pqcorrelation} for different activity values. In equilibrium, the above cross-correlation vanishes since the joint stationary distribution is factorized into the displacement and velocity parts, \cite{landau1980statistical}. In the NESS, however, \color{black} one finds 
\begin{align*}
    0 = \frac{\id}{\id t} \la q_i ; q_{i + \ell} \ra = \la v_i ; q_{i + \ell} \ra + \la q_i ; v_{i + \ell} \ra = \la q_{i + \ell} ; v_i \ra + \la q_i ; v_{i - \ell} \ra = C_{qv}(-\ell) + C_{qv}(\ell)
\end{align*}
\color{black}
so that $C_{qv}(\ell)$ is antisymmetric in $\ell$, 
%i.e., $C_{qv}(-\ell)=-C_{qv}(\ell)$ (
and equivalently, $C_{qv}(n-\ell)=-C_{qv}(\ell)$ due to periodic boundary conditions. \color{black} That also yields $C_{qv}(0) = C_{qv}(n/2) = 0$. \color{black} Additionally, it shows an oscillatory decay with $\ell$ (note that $\ell=n/2$ indicates the point of maximal separation for the periodic chain). The antisymmetric nature of the cross-correlations is also seen in a heat-conducting harmonic chain~\cite{rieder1967properties,dhar2001heat} coupled to reservoirs at different temperatures, where it is a signature of a steady heat current in the chain. Hence, it could indicate here as well the presence of a steady current in the NESS. As expected, this behavior is more prominent at higher activity.

\section{Conclusion}
\color{black} Coupling an oscillator chain to a medium of run-and-tumble particles is important for constructing and understanding fluctuating membranes in active media. Starting from the reciprocally coupled system-medium dynamics, \color{black} we have derived an explicit reduced Langevin dynamics for the chain, assuming a time-scale separation between the slow chain and the fast bath. The nonequilibrium nature of the medium manifests itself not only via a violation of the usual equilibrium FDR, but also through the emergence of a negative linear friction at high activity. To understand the fate of this linear instability, we investigated the late-time behavior numerically and found that the anti-damping is compensated for at nonlinear orders, similar to a Rayleigh oscillator-like behavior.  The corresponding nonequilibrium stationary condition is characterized by nontrivial fluctuations; the velocity and displacement distributions of the individual oscillators develop broad shoulders in the strongly active regime. At the collective level, the chain develops spatial organization, with both velocity and displacement correlations exhibiting damped oscillatory behavior. Moreover, the displacement–momentum cross-correlations are antisymmetric,  indicating a persistent steady current.

\begin{acknowledgments}
We thank Shin-ichi Sasa and Marco Baiesi for useful discussions. AB is supported by the Research Foundation - Flanders (FWO) doctoral fellowship 1152725N, and IS by the European Union’s Horizon 2024 research and innovation programme under the Marie Sklodowska–Curie (HORIZON-TMA-MSCA-PF-EF) grant agreement No.
101205210. \color{black} Finally, we are very grateful for anonymous reviews and comments. \color{black}
\end{acknowledgments}

\bibliographystyle{unsrt}
\bibliography{bib.bib}

\newpage

\begin{center}
{\LARGE \textbf{Appendix}}\\[1ex]
\end{center}
\appendix 
\section{Time scale separation}\label{appendix time scale separation}
We study the different scales in the forward generator $\cal L^\dagger$ of the full system, which appears in the Fokker-Planck equation for the dynamics Eq.~\eqref{p dot original}--\eqref{original z dynamics},
\begin{align*}
    &\frac{\partial \rho_{\text{tot}}}{\partial t}(\vec q, \vec p, \vec z,\vec \sigma) = \cal L^\dagger \rho_{\text{tot}}(\vec q, \vec p, \vec z,\vec \sigma) \nonumber \\
    & = \sum_{m = 1}^N - \frac{\partial}{\partial z_m}\left[\left( v_0 \sigma_m + \mu_zf_q(z_m)\right) \rho_{\text{tot}}(\vec q, \vec p, \vec z,\vec \sigma) \right] + \alpha [\rho_{\text{tot}}(q,p,\vec z,- \sigma_m) - \rho_{\text{tot}}(q,p,\vec z, \vec \sigma )]  \\
    & \qquad - \sum_{j=0}^{n-1} \frac{p_{j}}{M} \frac{\partial \rho_{\text{tot}}}{\partial q_{j}}(\vec q, \vec p, \vec z,\vec \sigma) + \left[\kappa (q_{j+1} + q_{j-1} - 2 q_{j})  - K q_{j} - \zeta  \,\sum_{m = 1}^N F(r_j-z_m)  \right] \frac{\partial \rho_{\text{tot}}}{\partial p_{j}}(\vec q, \vec p, \vec z,\vec \sigma) \nonumber
\end{align*}
where $f_q(z)=-\zeta\sum_{j=0}^{n-1}q_j\,\partial_z F(r_j-z)$. Let us assume that in the far past the chain is uncoupled from the active bath and that the $q_{j}$ oscillate with a typical amplitude $h$. That could be the initial amplitude of the oscillator string or the mean-square displacement $\sqrt{\langle q^2 \rangle}$ due to a surrounding temperature bath. From the uncoupled (\textit{e.g.} $\zeta = 0$) equation of motion Eq.~\eqref{p dot original}, one recognizes the typical timescale $t_{\text{chain}} = 1 /\omega_{\text{chain}}, \ \omega_{\text{chain}} = \sqrt{\max\{\kappa, K\}/M}$ and thus a typical momentum scale $p_{\text{chain}} = M \omega_{\text{chain}} h$. Similarly, from the uncoupled RTP dynamics, one recognises two possible timescales $t_{\text{RTP}} =  L/v_0$ or $t_{\text{RTP}} = \alpha^{-1}$. Time scale separation between a slow chain and fast bath then requires $t_{\text{RTP}} \ll t_{\text{chain}}$, which informally holds for large $v_0$ or large $\alpha$. In the current manuscript, we are interested in the nonequilibrium behaviour of the bath, while RTPs in the limit $\alpha \to \infty$ become equilibrium-like. As such, we take $t_{\text{RTP}} =  L/v_0$ as the relevant time scale for fixed $\alpha$.  
From dimensional analysis, the generator can then be written in the form $\cal L^\dagger = \cal L_{z, \sigma}^\dagger + \ve \left( \cal L_q^\dagger + \cal L_p^\dagger\right)$ with
\begin{align}\label{time scale separation epsilon}
    \ve = \frac{t_{\text{RTP}}}{t_{\text{chain}}} = \frac{\omega_{\text{chain}} L}{v_0} \ll 1 
\end{align}
and $\cal L_{z, \sigma}^\dagger, \cal L_q^\dagger, \cal L_p^\dagger$ the respective generators of the active particles, chain positions $q$ and mometa $p$ given by
\begin{align*}
      &\mathcal{L}^\dagger_{z, \sigma} \rho_{\text{tot}}  = \frac{v_0}{L} \sum_{m = 1}^N - \frac{\partial}{\partial(z_m/L)}\left[\left( \sigma_m + \frac{\mu_zf_q(z_m)}{v_0}\right) \rho_{\text{tot}} \right] + \alpha [\rho_{\text{tot}}(q,p,z,- \sigma_m) - \rho_{\text{tot}}(q,p,z, \sigma_m)] \\
      & \mathcal{L}^\dagger_{q} \rho_{\text{tot}} = \frac{v_0}{L} \sum_{j=0}^{n-1} \frac{p_{j}}{M \omega_{\text{chain}} h} \frac{\partial \rho_{\text{tot}}}{\partial (q_{j}/h)}     \\
       &\mathcal{L}^\dagger_{p} \rho_{\text{tot}}  = \frac{v_0}{L} \sum_{j=0}^{n-1}   \Bigg[ \frac{(q_{j+1} + q_{j-1} - 2 q_{j})}{h}  - \frac{K}{M\omega_{\text{chain}}^2} \frac{q_{j}}{h} - \zeta\, \frac{F_0}{M \omega_{\text{chain}}^2 h} \sum_{m = 1}^N \frac{F(r_j-z_m)}{F_0}  \Bigg] \frac{\partial \rho_{\text{tot}}}{\partial \left(\frac{p_{j}}{M \omega_{\text{chain}} h}\right)}
\end{align*}
By construction, we have, at least initially,  
\begin{align*}
     \frac{p_{j}}{M \omega_{\text{chain}} h} &= O(1), \qquad \frac{q_{j}}{h}  = O(1), \qquad  \frac{(q_{j+1} + q_{j-1} - 2 q_{j})}{h} = O(1), \qquad \frac{K}{M \omega_{\text{chain}}^2} = O(1)
\end{align*}
Moreover, we will assume that $F(x)/F_0 = O(1)$, meaning that the interaction force is bounded with typical amplitude $F_0$ (\textit{i.e.} it can't be a periodic Dirac delta function). Lastly, we take $\zeta$ small enough such that $\zeta \frac{F_0}{M \omega_{\text{chain}}^2 h} = O(1)$. Consequently, $\mathcal{L}_q^\dagger \sim \mathcal{L}_p^\dagger \sim \mathcal{L}^\dagger_{z, \sigma} \sim \frac{v_0}{L}$ for large $\frac{v_0}{L}$ such that $\mathcal{L}_q^\dagger/\mathcal{L}^\dagger_{z, \sigma} \sim \mathcal{L}_p^\dagger/\mathcal{L}^\dagger_{z, \sigma} \sim 1$ while $\ve \ll 1$ in $\cal L^\dagger = \cal L_{z, \sigma}^\dagger + \ve \left( \cal L_q^\dagger + \cal L_p^\dagger\right)$, leading to a time-scale separation in the system and a generator in the standard form to apply the projection operator technique. In other words, Eq.~\eqref{time scale separation epsilon} expresses that the typical time scale of the chain $1/\omega_{\text{chain}} = \sqrt{\frac{M}{\max\{\kappa, K\}}}$ should be much slower than that of the active particles, $L/v_0$ for large enough $v_0$ (at fixed $\alpha$). \\
When connecting the system to the bath, the time scale separation remains valid as long as $q \sim h$ but breaks down when $q \gg h$ dynamically (\textit{e.g.} due to the negative friction instability).

\section{Stationary distribution  \texorpdfstring{$\rho_q$}{ρq}}\label{appendix rho_q}
The solution to Eq.~\eqref{equation L rhoq = 0} factorizes due to the independence of the $z-$particles $\rho_q(\vec z, \vec \sigma) = \prod_{m = 1}^N \rho_q^m(z_m,\sigma_m)$ where $\rho_q^m$ solves
\begin{align}
  &0= -  \ \frac{\partial}{\partial z_m} \left( \left( v_0 \sigma_m + \mu_zf_q(z_m) \right) \rho_{q}^m(z_m, \sigma_m) \right) + \alpha \left(\rho_{q}^m(z_m, -\sigma_m)-  \rho_{q}^m(z_m, \sigma_m)\right) \nonumber
\end{align}
Its solution is known, \cite{beyen2025couplingelasticstringactive, Bena_2003}, which we write in the form
\begin{align}
\rho_q^m(z_m, \sigma_m) &= \frac{P_q(z_m) + \sigma_m Q_q(z_m)}{2}, \qquad  Q_q(z_m) = \frac{J_q - \mu_zf_q(z_m) \ P_q(z_m)}{v_0} \nonumber  \\
P_q(z_m) &= \frac{\psi(z_m)}{v_0^2 - \mu_z^2 f_q(z_m)^2}  \Bigg(C_2 - J_q \int_{0}^{z_m} \  \frac{(2 \alpha + \mu_zf'_q(x))}{\psi(x)} \ \id x \Bigg) \label{stationary p phi} \\
\psi(z_m) &= \exp{\Bigg[ \int_{0}^{z_m} \ I(y) \ \id y \Bigg]} > 0, \qquad I(y)  = 2 \alpha\frac{\mu_zf_q(y)}{v_0^2 - \mu_z^2 f_q(y)^2} \nonumber 
\end{align}
with $P_q(z_m) = \rho_q^m(z_m, \sigma_m) + \rho_q^m(z_m, -\sigma_m)$ the probability density that the probe is at location $z_m$ and $Q_q(z_m) = \sigma_m  \left[\rho_q^m(z_m, \sigma_m) - \rho_q^m(z_m, -\sigma_m) \right]$ the chirality. The constants $J_q, C_2$ follow from the periodic boundary conditions and normalization
\begin{align*}
    J_q &= C_2 \ \chi_q, \qquad \chi_q =  \frac{ \left(\psi(L) - 1 \right)}{\oint\left(2 \alpha+ \mu_zf'_q(x)\right) \frac{\psi(L)}{\psi(x)} \ \id x} \\
    C_2^{-1} &= \oint \frac{\psi(z)}{v_0^2- \mu_z^2 f_q(z)^2} \Bigg( 1 - \zeta \int_{0}^z \frac{\left(2 \alpha+\mu_zf'_q(x)\right)}{\psi(x)} \, \  \id x \, \Bigg) \ \id z 
\end{align*}
Note in particular that $\rho_q(z, \sigma)$ does \textit{not} have the Boltzmann form $e^{- \beta H_q}$ as we are out-of-equilibrium.  \\

For small coupling $\zeta \ll 1$, writing $f_q(z) = \zeta \tilde{f}_q(z)$, the functions $P_q$ and $Q_q$ simplify to
\begin{align}
            P_q(z_m) 
            &= \frac{1}{L} \Bigg[1 + \zeta \beta_{\text{eff}} \Bigg( \int_{0}^{z_m} \tilde{f}_q(y) \ \id y + \left(\frac{L}{2} - z_m \right) \frac{1}{L} \oint \id x \ \tilde{f}_q(x)  - \frac{1}{L} \oint \id z  \int_{0}^{z} \tilde{f}_q(y) \ \id y\Bigg)\Bigg]  + O(\zeta^2)  \nonumber \\ 
             Q_q(z_m)  &=  -\zeta \frac{\mu_z}{L v_0}\left(\tilde{f}_q(z_m)  -  \frac{1}{L} \oint \id x \ \tilde{f}_q(x)\right) + O(\zeta^2) \nonumber 
        \end{align}
The leading order $O(\zeta^0)$ gives the uniform distribution $P_q^{m}(z) \Big|_{\zeta = 0} = 1/L$, as expected for the stationary distribution of a collection of free, independent, active particles on the ring. Computing the integrals yields $ \oint \id x  \ \tilde{f}_q(x) = 0$ and
\begin{align*}
    \int_{0}^z \tilde{f}_q(y) \ \id y 
    & = -   \sum_{\ell = 0}^{n-1} q_{\ell } \left(F(r_\ell  - z) - F(r_\ell ) \right) \\
     \frac{1}{L} \oint \id z  \int_{0}^z \tilde{f}_q(y) \ \id y 
       &= -    \sum_{\ell = 0}^{n-1} q_{\ell } \left(\frac{1}{L} \oint \id x \ F(x) - F(r_\ell)  \right)
\end{align*}
In the last line, we used the fact that for a periodic function $F(x) = F(x + L)$
\begin{align}
&\forall \  a \in \mathbb{R}: \qquad    \int_a^{a+L} \id x \ F(x) = \oint \id x \ F(x)  \label{periodic integral 1}\\
   & \frac{1}{L} \oint \id z \  F(r_\ell  - z) = \frac{1}{L} \int_{\ell  - L}^{\ell } \id x \ F(x) = \frac{1}{L} \oint \id x \ F(x) =   \frac{1}{L} \oint \id x \ F(x) \label{periodic integral 2}
\end{align}
Hence,
\begin{align}\label{rho phi small coupling}
     P_q(z_m) & 
            = \frac{1}{L} \Bigg[1 - \zeta  \beta_{\text{eff}} \sum_{\ell = 0}^{n-1} q_{\ell } \left(F(r_\ell  - z_m) - \frac{1}{L} \oint \id x \ F(x) \right)   \Bigg]  + O(\zeta^2) \\
            Q_q(z_m) &= \zeta   \frac{\mu_z}{L v_0} \sum_{\ell = 0}^{n-1} q_{\ell } \partial_{z_m} F(r_\ell  - z_m) + O(\zeta^2) \nonumber
\end{align}

\section{Calculation of the reduced dynamics}\label{appendix calculation}
We add details on how to compute the induced streaming term, friction coefficient, and noise amplitude.
 
\subsection{Streaming term}\label{appendix streaming term}

In this section, we compute the streaming term $\bar{F}(q)$ in Eq.~\eqref{reduced dynamics 1}:
\begin{align}
    \zeta \bar{F}(q) & = \zeta N \oint \id z \ F(r_j - z)  \sum_{\sigma = \pm 1} \rho_q(z,\sigma) 
    = \zeta N \oint \id z \ F(r_j - z) \ P_q(z) \label{average v_0}
\end{align}
Equation \eqref{stationary p phi} can be substituted into Eq.~\eqref{average v_0}, but the resulting integrals do not reduce to a simple or manageable form. Instead, for weak coupling $\zeta\ll 1$, we use the form Eq.~\eqref{rho phi small coupling}
\begin{align*}
    \zeta \bar{F}(q) &= \zeta \frac{N}{L} \oint  \id x \ F(x) - \zeta^2 N\beta_{\text{eff}} \sum_{\ell = 0}^{n-1} \frac{1}{L} \oint \id z \ F(r_j-z) \left(F(r_\ell  - z) - \frac{1}{L} \oint \id x \ F(x) \right)  \  q_\ell + O(\zeta^3)
\end{align*}
The term in front of $q_\ell$ only depends on $r_j-r_\ell$ since substituting $y = r_\ell-z$ yields
\begin{align*}
   &\frac{1}{L} \oint \id z \ F(r_j-z) \left(F(r_\ell  - z) - \frac{1}{L} \oint \id x \ F(x) \right)  \\
   &=  \frac{1}{L} \int_{r_\ell-L}^{r_\ell} \id y \ F(r_j - r_\ell + y) \left(F(y) - \frac{1}{L} \oint \id x \ F(x) \right) \\
    & = \frac{1}{L} \oint \id y \ F(r_j - r_\ell + y) \left(F(y) - \frac{1}{L} \oint \id x \ F(x) \right) 
\end{align*}
where we have used Eq.~\eqref{periodic integral 1}. Therefore,
\begin{align*}
\zeta \bar{F}(q)
&=
\zeta \frac{N}{L} \oint  \id x \ F(x)
-\zeta^2 N \beta_{\text{eff}}
\sum_{\ell=0}^{n-1}{\cal M}_{r_j-r_\ell}q_\ell
+O(\zeta^3) \\
{\cal M}_r
&= \frac{1}{L} \oint \id x \  F(r+x) \ F(x) -\left(\frac{1}{L} \oint \id x \  F(x) \right)^2 
\end{align*}
which is Eq.~\eqref{streaming weak coupling}. For the von Mises form Eq.~\eqref{von mises}, one finds
\begin{align*}
    &\frac{1}{L} \oint  \id x \ F(x) = F_0, \qquad \cal M_r = 2 F_0^2 \sum_{k = 1}^\infty \frac{I_k(p)^2}{I_0(p)^2} \cos\left(\frac{2 \pi k}{L} r \right) = F_0^2 \left(\frac{I_0\left(2 p \cos\left(\frac{\pi r}{L} \right) \right)}{I_0(p)^2}  - 1 \right)
\end{align*}
where we have used the properties, \cite{watson1995treatise},
\begin{align*}
    I_0(p) &= \frac{1}{2\pi} \int_0^{2 \pi} e^{p \cos \theta} \ \id \theta = \frac{1}{L} \oint \id x \ e^{p \cos\left(\frac{2 \pi }{L} x \right)} \\
    \sum_{k = - \infty}^\infty I_k(p)^2 e^{i k \theta} &= I_0(2 p \cos(\theta/2))
\end{align*}
Going to Fourier space, following Eq.~\eqref{dft transform}, we have 
\begin{align}
    &\kappa (q_{j+1} + q_{j-1} - 2 q_{j}) - K q_j = \sum_{k = 0}^{n-1} - \left[2\kappa \left(1-\cos \left(\frac{2 \pi k}{n} \right)\right) + K \right]\tilde{q}_{k} \ e^{2 \pi i k j/n} \label{dft step 1} \\
    &\zeta \bar{F}(q) =  \sum_{k = 0}^{n-1} \left[\zeta \frac{N}{L} \oint  \id x \ F(x) \delta_{k,0}
- \left(\zeta^2 N \beta_{\text{eff}}
\sum_{\ell=0}^{n-1}{\cal M}_{r_j-r_\ell} e^{-2 \pi i k (j-\ell) /n} \right)  \tilde{q}_k \right] e^{2 \pi i k j /n} \label{dft step 2} \\
&M\frac{\id^2 q_j}{\id t^2}  = \sum_{k = 0}^{n-1} M \frac{\id^2 \tilde{q}_k}{\id t^2} \ e^{2 \pi i k j/n} \label{dft step 3}
\end{align}
Moreover, since $F$ is periodic, it can be expanded in a Fourier series, yielding
\begin{align}
    &F(x) = \sum_{m \in \mathbb{Z}} \tilde{F}_m \ e^{2 \pi i m x/L} 
    , \qquad  \frac{1}{L} \oint \id x \ F(x) = \tilde{F}_0\label{discrete fourier series F(x)}, \qquad \cal M_r  = \sum_{m \in \mathbb{Z}} (1 - \delta_{m,0}) \  |\tilde{F}_m|^2 \ e^{2 \pi i mr/L}
\end{align}
Consequently, for $k \in  \{0,..,n-1\}$,
\begin{align*}
     \sum_{\ell=0}^{n-1}{\cal M}_{r_j-r_\ell} e^{-2 \pi i k (j-\ell) /n} = \sum_{m \in \mathbb{Z}} ( 1 - \delta_{m,0}) |\tilde{F}_m|^2  \sum_{\ell=0}^{n-1} e^{2 \pi i (m-k)(j-\ell)/n}=n \left(1-\delta_{k,0}\right) \sum_{\ell \in \mathbb Z} |\tilde{F}_{k + \ell n}|^2     \nonumber 
\end{align*}
which follows from the identities
\begin{align*}
   \frac{1}{L} \oint \id x \  e^{2 \pi i k x/L}  = \delta_{k,0}, \qquad \frac{1}{n} \sum_{\ell = 0}^{n-1} e^{2 \pi i (m-k) \ell/n}  =\sum_{\ell \in \mathbb{Z}} \delta_{m, k + \ell n}
\end{align*}
Since the only $j$-dependence in Eqs.~\eqref{dft step 1}--\eqref{dft step 3} is in the $e^{2 \pi i k j/n}$, the equations of motion for \newline  $k \in  \{0,..,n-1\}$ are equivalent to 
\begin{align}
     M \frac{\id^2 \tilde{q}_k}{\id t^2} & = -\left[2\kappa \left(1 - \cos \left(\frac{2 \pi k}{n} \right)\right) + K_{\text{eff},k}\right] \tilde{q}_k -  \zeta N \tilde{F}_0\delta_{k,0} \nonumber \\
     K_{\text{eff},k}      & =K - \left( 1-\delta_{k ,0} \right) \zeta^2  N n \beta_{\text{eff}}  \sum_{\ell \in \mathbb{Z}}  |\tilde{F}_{k + \ell n}|^2 \nonumber
    \end{align}
    which is Eq.~\eqref{fourier space mean force only}. \color{black} The last term only acts on the zeroth mode $k = 0$ for which we find
    \begin{align}
     M \frac{\id^2 \tilde{q}_0}{\id t^2} + K \tilde q_0 & = -  \zeta N \tilde{F}_0 \nonumber
    \end{align}
    with solution
    \begin{align*}
        \tilde q_0(t) = - \frac{\zeta N \tilde F_0}{K} + \left(\tilde q_0(0) + \frac{\zeta N \tilde F_0}{K} \right) \cos\left(\sqrt{\frac{K}{M}} t \right) + \frac{\frac{\id \tilde q}{\id t}(0)}{\sqrt{\frac{K}{M}}} \sin\left(\sqrt{\frac{K}{M}} t \right)
    \end{align*}
    with a downward shift given by $- \frac{\zeta N \tilde F_0}{K}$. 
    \color{black}

\subsection{Noise term}\label{appendix noise term}
Explicitly writing out the covariance Eq.~\eqref{def B full} yields
\begin{align}
     B_{j \ell} &=\zeta^2 N \int_0^{\infty} \id \tau \Bigg[ \oint \id z  \  \id z_{0} \  F(r_j-z)  \ F(r_\ell  - z_{0}) \sum_{\sigma, \sigma_{0} = \pm 1} \rho_q(z, \sigma,\tau|z_{0}, \sigma_{0}) \   \rho_q(z_{0}, \sigma_{0})  \nonumber \\
    &    \hspace{8 cm}   - \left( \oint \id z \ F(r_j-z) \sum_{\sigma} \rho_q(z, \sigma)  \right)^2 \Bigg] \nonumber
\end{align}
Here, $\rho_q(z, \sigma,\tau|z_{0}, \sigma_{0})$ is the transition probability  which is calculated in \cite{beyen2025couplingelasticstringactive} to leading order $O(\zeta^0)$,
\begin{align}
    \rho_q(z, \sigma,\tau|z_{0}, \sigma_{0}) = \frac{e^{- \alpha \tau}}{L} \sum_{n \in \mathbb{Z}} &e^{i2 \pi n(z-z_{0})/L} \Bigg\{\left(\frac{1 + \sigma \sigma_{0}}{2} \right) \cosh(\Upsilon_n \tau) \label{transition prob}\\
    & +  \left[\alpha \left(\frac{1 - \sigma_m \sigma_{m,0}}{2} \right) - \frac{2 \pi i n v_0}{L} \left(\frac{\sigma + \sigma_{0}}{2} \right) \right] \frac{\sinh(\Upsilon_n \tau)}{\Upsilon_n}  \Bigg\} \nonumber
\end{align}
with $ \Upsilon_n  = \sqrt{\alpha^2 -\frac{4 \pi^2n^2}{L^2} v_0^2}$. 
Expanding $F(x)$ using Eq.~\eqref{discrete fourier series F(x)} and $r_j = j L/n$, the noise term becomes a Fourier series
\begin{align*}
     B_{j \ell}  = \sum_{a,c \in \mathbb Z} b_{ac} \ e^{2 \pi i aj/n} e^{2 \pi i c \ell/n}
\end{align*}
where
\begin{align*}
  b_{ac}  &=\zeta^2 \tilde{F}_a \tilde{F}_c N \int_0^{\infty} \id \tau \Bigg[ \oint \id z  \ \id z_{0} \  e^{-2 \pi i a z/L} \ e^{- 2 \pi i c z_0/L}  \sum_{\sigma, \sigma_{0} = \pm 1}   \rho_q(z, \sigma,\tau|z_{0}, \sigma_{0}) \rho_q(z_{0}, \sigma_{0}) \nonumber \\
    & \hspace{8 cm}   - \left( \oint \id z \ e^{-2 \pi i a z/L} \sum_{\sigma} \rho_q(z, \sigma)  \right)^2 \Bigg]
\end{align*}
The calculation of $b_{ac}$ can be found in the supplementary material of \cite{beyen2025couplingelasticstringactive} leading to
\begin{align*}
   b_{ac} =\delta_{c,-a} (1 -\delta_{a,0}) \ \zeta^2 \frac{N L^2 \alpha}{2 \pi^2 v_0^2 a^2} |\tilde{F}_a|^2
\end{align*}
and thus
\begin{align*}
    B_{j \ell} & = \zeta^2 \frac{N L^2 \alpha}{2 \pi^2 v_0^2} \sum_{a \in \mathbb Z} \frac{(1 -\delta_{a,0})}{a^2}   |\tilde{F}_a|^2 \ e^{2 \pi ia (j- \ell)/n } = \zeta^2 \frac{N L^2 \alpha}{ \pi^2 v_0^2} \sum_{a = 1}^{\infty} \frac{|\tilde{F}_a|^2}{a^2}  \cos \left( \frac{2 \pi a}{n} (j - \ell) \right) 
\end{align*}
which is Eq.~\eqref{friction and noise}. In Fourier space, the noise term satisfies
\begin{align}\label{dft transform noise}
\xi_j(t)&=\sum_{k=0}^{n-1}\tilde \xi_k(t)e^{2\pi i k j/n},
\qquad
\tilde \xi_k(t)=\frac{1}{n}\sum_{j=0}^{n-1}\xi_j(t)e^{-2\pi i k j/n}, \qquad \left \langle \tilde \xi_k(t) \right \rangle  = 0
\end{align}
and for $k,m \in \{0,...,n-1\}$
\begin{align}
\left \langle \tilde \xi_k(t) \ \tilde \xi_m(t')^* \right \rangle &= \frac{2\delta(t-t')}{n^2} \sum_{j=0}^{n-1} \sum_{\ell =0}^{n-1} B_{j \ell} e^{-2\pi i k j/n}  e^{2\pi i m \ell/n} \nonumber \\
& =\delta(t-t') \  \zeta^2 \frac{2N L^2 \alpha}{2 \pi^2 n^2 v_0^2} \sum_{a \in \mathbb Z} \frac{(1 -\delta_{a,0})}{a^2}   |\tilde{F}_a|^2  \sum_{j=0}^{n-1} \sum_{\ell =0}^{n-1}  e^{2 \pi ia (j- \ell)/n } e^{-2\pi i k j/n}  e^{2\pi i m \ell/n} \nonumber\\
& =\delta(t-t') \  (1 -\delta_{k,0}) \ \delta_{k,m} \zeta^2 \frac{2N L^2 \alpha}{2 \pi^2 v_0^2} \sum_{a \in \mathbb Z}\frac{|\tilde{F}_{k + a n}|^2}{(k+an)^2}  \nonumber \\
& = 2 \tilde{B}_{k} \ \delta_{k,m} \delta(t-t')
\end{align}
\subsection{Friction term}\label{appendix friction term}
The covariance $\langle \cdot \ ; \ \cdot \rangle^{\text{BO}}_q$ in the formula Eq.~\eqref{def friction coefficient} for $\nu_{j \ell}$ can be rewritten as a single expectation value
\begin{align}
    \nu_{j \ell}(q) &=  -\zeta N \int_0^\infty \id\tau\,
\left\langle
F(r_j-z(\tau)) \cdot 
\frac{\partial \log \rho_q}{\partial q_\ell}(z, \sigma)
\right\rangle_q^{\text{BO}}\\
    & \qquad + \zeta N \int_0^\infty \id\tau\,
\left\langle
F(r_j-z(\tau)) \right\rangle_q^{\text{BO}} \cdot \left \langle
\frac{\partial \log \rho_q}{\partial q_\ell}(z, \sigma)
\right\rangle_q^{\text{BO}} \nonumber \\
    & =  -\zeta N \int_0^\infty \id\tau\,
\left\langle
F(r_j-z(\tau)) \cdot 
\frac{\partial \log \rho_q}{\partial q_\ell}(z, \sigma)
\right\rangle_q^{\text{BO}}\label{reduced form nu 2}
\end{align}
where we have used the normalization of $\rho_q$ to eliminate the second term since
\begin{align*}
    \left \langle  \frac{\partial \log \rho_q}{\partial q_\ell}(\sigma)   \right \rangle_q^{\text{BO}} & = \sum_{\sigma = \pm 1} \oint \id z \ \frac{\partial \rho_q}{\partial q_\ell}(\sigma) =  0
\end{align*}
Writing out the last expression in Eq.~\eqref{reduced form nu 2} yields the form
Explicitly writing out $\nu$
\begin{align*}
    \nu_{j \ell}  & = - \zeta N\int_0^{\infty} \id \tau \oint \id z \ \id z_{0} \ F(r_j-z)  \sum_{\sigma, \sigma_{0} = \pm 1} \rho_q(z, \sigma,\tau|z_{0}, \sigma_{0})   \ \frac{\partial \rho_q}{\partial q_{\ell }} (z_{0}, \sigma_{0})
\end{align*}
We first compute the derivative
\begin{align*}
      \frac{\partial \rho_{q}}{\partial q_{\ell }}(z) &=   \frac{\partial} {\partial q_{\ell }} \left( \frac{P_q(z) + \sigma Q_q(z)}{2} \right)  = - \frac{\zeta  \mu_z}{2 L v_0} \left[ \frac{2 \alpha}{v_0} \left(F(r_\ell  - z) - \frac{1}{L} \oint  \id x \ F(x) \right) - \sigma \partial_{z} F(r_\ell  - z)\right] \\
      & = - \frac{\zeta  \mu_z}{2 L v_0} \sum_{k = - \infty}^{\infty}  \left[ \frac{2 \alpha}{v_0}  + \frac{2 \pi i k \sigma}{L} \right] (1 -\delta_{k,0}) \ \tilde{F}_k \ e^{2 \pi i k(r_\ell  - z)/L}
\end{align*}
The friction can then be written in a Fourier series
\begin{align*}
     \nu_{j \ell}  = \sum_{a,b \in \mathbb Z} v_{a b} \ e^{2 \pi i a j/n} e^{2 \pi i b \ell/n}
\end{align*}
where
\begin{align*}
   v_{a b} =  \frac{\zeta^2 N\mu_z (1 -\delta_{b,0})}{2 L v_0} \tilde{F}_a \tilde{F}_b  \int_0^{\infty} \id \tau &\oint \id z \ \id z_{0} \ e^{-2 \pi i a z/L} \  e^{- 2 \pi i b z/L}   \sum_{\sigma, \sigma_{0 = \pm 1}}  \rho_q(z, \sigma,\tau|z_{0}, \sigma_{0})  \ \left[ \frac{2 \alpha}{v_0}  + \frac{2 \pi i k \sigma}{L} \right]
\end{align*}
with the same transition probability as in Eq.~\eqref{transition prob}. The calculation of $v_{ab}$ can be found in the supplementary material of \cite{beyen2025couplingelasticstringactive} leading to
\begin{align*}
    v_{a b} =  \frac{\zeta^2 N\mu_z}{v_0^2} \tilde{F}_a \tilde{F}_b (1 -\delta_{b,0}) \  \delta_{a,-b} \ \left(\frac{\alpha^2 L^2}{ \pi^2 v_0^2 a^2} -1\right)
\end{align*}
and thus
\begin{align*}
     \nu_{j \ell} &= \frac{\zeta^2 N\mu_z}{v_0^2} \sum_{a\in \mathbb Z}(1-\delta_{a,0}) \  |\tilde{F}_a|^2  \left(\frac{\alpha^2 L^2}{ \pi^2 v_0^2 a^2} -1\right) \ e^{2 \pi ia (j- \ell)/n} \\
     & = \frac{2\zeta^2 N\mu_z}{v_0^2} \sum_{a = 1}^\infty \  |\tilde{F}_a|^2  \left(\frac{\alpha^2 L^2}{ \pi^2 v_0^2 a^2} -1\right) \ \cos \left(\frac{2 \pi a}{n}(j - \ell)  \right)
\end{align*}
which is Eq.~\eqref{friction and noise}. Lastly, in Fourier space, the friction term then becomes
\begin{align*}
    \sum_{\ell=0}^{n-1}\nu_{j\ell}\frac{\id q_\ell}{\id t} & = \frac{\zeta^2 N\mu_z}{v_0^2} \sum_{a \in \mathbb{Z}}(1-\delta_{a,0}) \  |\tilde{F}_a|^2  \left(\frac{\alpha^2 L^2}{ \pi^2 v_0^2 a^2} -1\right)  \sum_{\ell=0}^{n-1} \sum_{k=0}^{n-1}\frac{\id \tilde q_k}{\id t}e^{2\pi i k \ell/n} e^{2 \pi ia(j- \ell)/n} \\
& = \sum_{k=0}^{n-1} (1-\delta_{k,0})\frac{\zeta^2 Nn\mu_z}{v_0^2} \sum_{a \in \mathbb{Z}} \  |\tilde{F}_{k + a n}|^2  \left(\frac{\alpha^2 L^2}{ \pi^2 v_0^2 (k + a n)^2} -1\right)   \frac{\id \tilde q_k}{\id t}e^{ \frac{2 \pi ik}{n} j} \\
 & = \sum_{k=0}^{n-1} \tilde \nu_k   \frac{\id \tilde q_k}{\id t}e^{ \frac{2 \pi ik}{n} j}
\end{align*}

\subsection{Langevin equation}\label{appendix langevin equation}
Finally, the reduced dynamics in Fourier space becomes
\begin{align}
    & M\frac{\id^2 \tilde{q}_k}{\id t^2} + \tilde \nu_k\frac{\id \tilde{q}_k}{\id t} + M\omega_{k}^2 \tilde{q}_k = -\zeta \delta_{k,0} N \tilde{F}_0 + \tilde{\xi}_{k} \label{second fourier eq appendix}\\
    & \left \langle \tilde{\xi}_{k} \right \rangle = 0, \qquad \left \langle \tilde{\xi}_{k}(t) \ \tilde{\xi}_{m}(t') \right \rangle =  2 \tilde{B}_{k} \ \delta_{k,m} \delta(t-t') \nonumber \\ 
    &\tilde{\nu}_k= \frac{n}{2}(1-\delta_{k,0})\sum_{a \in \mathbb{Z}} \cal V_{k + a n}, \qquad \tilde{B}_{k} = \frac{(1 -\delta_{k,0}) }{2} \sum_{a \in \mathbb Z} \cal B_{k + a n} \nonumber 
\end{align}
with $\omega_k$ from Eq.~\eqref{omegak} and $\cal V_a, \cal B_a$ from Eq.~\eqref{V a  B a}. Consequently, each mode behaves independently as an (anti)-damped harmonic oscillator with noise, whose statistics are given by
\begin{align}
    \langle \tilde{q}_k(t) \rangle &=  e^{- \frac{\tilde \nu_k}{2 M}  t} \left[ \tilde{q}_k(0) \cos\left(\Omega_k t \right) + \frac{\tilde \nu_k \tilde{q}_k(0) + 2 M \tilde{q}_k'(0)}{2 M \Omega_k} \sin\left(\Omega_k t \right) \right]  \nonumber \\
      & \qquad + \delta_{k,0} \frac{\zeta N\tilde{F}_0 }{ M \omega_k^2} \left[\frac{\tilde \nu_k}{2 M \Omega_k} \sin(\Omega_k t) e^{- \frac{\tilde \nu_k}{2 M}  t} - \left(1-\cos(\Omega_k t) e^{- \frac{\tilde \nu_k}{2 M}  t} \right)   \right]\nonumber \\
    \text{Var}\left(\tilde{q}_k(t) \right)_\xi & = \langle \tilde{q}_k(t) \ \tilde{q}_k^*(t) \rangle - \langle \tilde{q}_k(t) \rangle \ \langle \tilde{q}_k^*(t) \rangle \nonumber \\
      & = \frac{\tilde B_k}{M\omega_k^2\tilde \nu_k} \left( 1-e^{-\frac{\tilde  \nu _k}{M}t} -\frac{\tilde\nu _k}{2M \Omega_k} \sin \left(2  \Omega _kt\right) e^{-\frac{ \tilde \nu _k}{M}t} -\frac{\tilde \nu_k^2}{2M^2 \Omega_k^2} \sin ^2\left( \Omega _k t\right)e^{-\frac{\tilde  \nu _k}{M}t} \right) \nonumber
 \end{align}
where $  \Omega_k^2 = \omega_k^2 - \frac{\tilde \nu_k^2}{4 M^2}$ and we have assumed the mode $k$ to be underdamped, $\tilde \nu_k <  2 M \omega_k$, being consistent with the order in the coupling where the friction is second order in $\zeta$ while the frequency is $O(\zeta^0)$. We thus recognise two possibilities for the late-time behaviour
\begin{itemize}
    \item If $\tilde \nu_k > 0$, then there is an exponential decay, leading to a constant variance
    \begin{align}
       \lim_{t \to \infty} \text{Var}\left(\tilde{q}_k(t) \right) = \frac{\tilde B_k}{M \omega_k^2\tilde \nu_k} \label{fluctuations phi n at late times appendix}
    \end{align}
and the chain keeps fluctuating. In equilibrium, we have $\tilde \nu_k = \beta \tilde B_k$ and we recover the equipartition result $\lim_{t \to \infty} \text{Var}\left(\tilde{q}_k(t) \right)_\xi = k_B T/(M \omega_k^2)$.
        \item  If $\tilde \nu_k < 0$, then the terms grow exponentially, indicating an instability in the system.
\end{itemize}

\section{Numerical implementation}
For the numerical results, we simulate equations Eq.~\eqref{eq:langevins0} using an Euler integration method with a time step typically in the range $\Delta t\sim O( 10^{-5})$. The transient fluctuations (in Fig.~\ref{fig:msdmsv}) are obtained by averaging trajectories starting from $t=0$ till $t=10^5$. For the stationary fluctuations, on the other hand, we first relax the particle to a stationary state for $t\sim O(10^5)$, and then do a time averaging in the stationary state.

\end{document}